\documentclass[12pt]{article}

\usepackage{multirow}
\usepackage{comment}
\usepackage{epsfig}
\usepackage{amssymb,amsmath}

\usepackage{setspace}

\newcommand{\bea}{\begin{eqnarray}}
\newcommand{\eea}{\end{eqnarray}}

\numberwithin{equation}{section}

\begin{document}
\begin{titlepage}
%
%
\vspace*{10mm}
\begin{center}
\baselineskip 25pt 
{\Large\bf
Inflation, Proton Decay, and Higgs-Portal Dark Matter in $SO(10) \times U(1)_\psi$
}
\end{center}
\vspace{5mm}
\begin{center}
{\large
Nobuchika Okada\footnote{okadan@ua.edu}, 
Digesh Raut\footnote{draut@udel.edu}, 
and Qaisar Shafi\footnote{shafi@udel.edu}
}
\end{center}
\vspace{2mm}

\begin{center}
{\it
$^{1}$ Department of Physics and Astronomy, \\ 
University of Alabama, Tuscaloosa, Alabama 35487, USA \\
$^{2,3}$ Bartol Research Institute, Department of Physics and Astronomy, \\
 University of Delaware, Newark DE 19716, USA
}
\end{center}
\vspace{0.5cm}
\begin{abstract}
We propose a simple non-supersymmetric grand unified theory (GUT) based on the gauge group $SO(10) \times U(1)_\psi$. 
The model includes 3 generations of fermions in ${\bf 16}$ ($+1$), ${\bf 10}$ ($-2$) and ${\bf 1}$ ($+4$) representations. 
The ${\bf 16}$-plets contain Standard Model (SM) fermions plus right-handed neutrinos, and the ${\bf 10}$-plet and the singlet fermions are introduced to make the model anomaly-free. 
Gauge coupling unification at $M_{GUT} \simeq 5 \times 10^{15}-10^{16}$ GeV is achieved by including an intermediate Pati-Salam breaking at $M_{I} \simeq  10^{12}-10^{11}$ GeV, which is a natural scale for the seesaw mechanism. 
For $M_{I} \simeq 10^{12}-10^{11}$, proton decay will be tested by the Hyper-Kamiokande experiment.     
The extra fermions acquire their masses from $U(1)_\psi$ symmetry breaking, and a $U(1)_\psi$ Higgs field drives a successful inflection-point inflation with a low Hubble parameter during inflation, $H_{inf} \ll M_{I}$. 
Hence, cosmologically dangerous monopoles produced from $SO(10)$ and PS breakings are diluted away. 
The reheating temperature after inflation can be high enough for successful leptogenesis. 
With the Higgs field contents of our model, a ${\bf Z}_2$ symmetry remains unbroken after GUT symmetry breaking, and the lightest mass eigenstate among linear combinations of the ${\bf 10}$-plet and the singlet fermions serves as a Higgs-portal dark matter (DM). 
We identify the parameter regions to reproduce the observed DM relic density while satisfying the current constraint from the direct DM detection experiments. 
The present allowed region will be fully covered by the future direct detection experiments such as LUX-ZEPLIN DM experiment. 
In the presence of the extra fermions, the SM  Higgs potential is stabilized up to $M_{I}$. 

\end{abstract}
\end{titlepage}

\section{Introduction}
\label{sec:Intro}
The lure of Grand Unified Theories (GUTs) is that the Standard Model (SM) gauge symmetry, $SU(3)_c \times SU(2)_L \times U(1)_Y$, is unified into a single gauge group, so that the three SM gauge interactions originate from a single theory.       
Accordingly, the SM quarks and leptons are unified into certain representations of the GUT gauge group, leading to the quantization of their electric charges \cite{GUT}.  
Supersymmetric (SUSY) GUT models have been  commonly studied in the literature, motivated by the fact that three SM gauge couplings are successfully unified at the GUT scale $M_{GUT} \simeq 10^{16}$ GeV with the weak scale SUSY \cite{SUSYGUT}.  
However, there is no evidence of the weak scale SUSY in the current data of the Large Hadron Collider experiments. 
This fact drives a renewed interest of non-SUSY GUTs in recent years.

Among GUT models, an $SO(10)$ framework is arguably one of the most appealing scenario \cite{nonSUSYGUT}, where the SM fermions in each generation are nicely unified into a single ${\bf 16}$ representation of the $SO(10)$ gauge group along with a SM singlet right-handed neutrino (RHN). 
In the non-SUSY $SO(10)$ GUT framework, we may consider the spontaneous symmetry breaking (SSB) of $SO(10)$ in two steps down to the SM gauge groups \cite{S010nonSUSY1, S010nonSUSY}: 
For example, the $SO(10)$ group is first broken down to the Pati-Salam (PS) group $SU(4)_c \times SU(2)_L\times SU(2)_R$ at $M_{GUT}\simeq 10^{16}$ GeV.  
Next, the PS gauge group is broken to the SM gauge group $SU(3)_c \times SU(2)_L \times U(1)_Y$ at an intermediate scale $M_I \simeq 10^{11}$ GeV. 
Associated with the PS SSB, the Majorana masses for the RHNs are generated, which play the key role in the seesaw mechanism \cite{Seesaw} for generating light SM neutrino masses. 
The mass scale of RHNs at the intermediate scale is a natural scale for the seesaw mechanism.  
Leptogenesis \cite{leptogenesis1} is a very simple mechanism to generate the observed baryon asymmetry 
   through the CP-violating out-of-equilibrium decay of Majorana RHNs. 
This scenario is automatically implemented in the $SO(10)$ GUT framework.    
Using a minimal set of Higgs fields, one ${\bf 10}$-plet and one ${\bf 126}$-plet, 
   realistic fermion mass matrices can be reproduced (see, for example, Ref.~\cite{S010nonSUSY}).

In general, GUT SSB produces stable topological defects such as monopoles and strings \cite{Tdefects, monopole1, monopole2, monopole3}.  
In the above example of two-step $SO(10)$ breaking, both the $SO(10)$ and the PS SSBs produce monopoles with their masses 
 of order of the SSB scales \cite{monopole2}. 
Since such super-heavy monopoles would be over-abundant before the Big Bang Nucleosynthesis \cite{monopole3}, a mechanism to significantly reduce the monopole density is necessary for reproducing our universe. 
One of the original motivation of the cosmological inflation scenario was to solve this monopole problem by diluting the monopole density \cite{Guth}. 
To sufficiently dilute the monopoles, the inflation must take place after the SSB, or equivalently, the Hubble parameter during the inflation ($H_{inf}$) must be smaller than the SSB scale. 
For well-known simple inflation scenarios, such as an inflation with a Coleman-Weinberg type potential \cite{Shafi:2006cs} and quartic inflation with non-minimal gravitational coupling, we estimate $H_{inf} \simeq 10^{13-14}$ GeV \cite{Okada:2010jf, Okada:2014lxa}. 
Although such inflationary scenarios can inflate away the GUT scale monopoles, the intermediate scale monopoles still survive if $M_I < H_{inf}$ \cite{Senoguz:2015lba}. 
Hence, we need a ``low-scale inflation scenario'' with  $H_{inf} < M_I$ to dilute the intermediate-sale monopoles.

Hybrid inflation \cite{Linde:1993cn} is a well-known example of  low-scale inflation scenario where the introduction of multi-scalar fields is crucial for realizing  inflation. 
Another interesting example is the so-called inflection-point inflation (IPI) scenario which can be realized with a single scalar field. 
In IPI, the inflaton potential exhibits an approximate inflection-point and slow-roll inflation occurs in the vicinity of the inflection-point.  
In Ref.~\cite{IPI}, a successful IPI scenario has been proposed in the context of a $U(1)$ Higgs-Yukawa model where the  Higgs field is identified with the inflaton field. 
In the model, the renormalization group (RG) improved effective potential of the inflaton/Higgs field realizes an approximate inflection-point at a scale $M$ if the running inflaton/Higgs quartic coupling $\lambda$ exhibits a minimum with almost vanishing value at $M$, namely 
$\lambda (\phi =M) \simeq  0$ and its beta-function $\beta_\lambda (\phi = M)\simeq 0$. 
To satisfy these conditions, it is crucial for the inflaton field to have both gauge and Yukawa interactions, and the gauge and Yukawa couplings at $\phi = M$ must be balanced to achieve $\beta_\lambda (\phi = M)\simeq 0$.
A successful IPI scenario in Ref.~\cite{IPI} leads to an upper bound, $H_{\rm inf} \lesssim 10^{10}$ GeV.

In this paper, we propose a simple non-SUSY GUT model based on the gauge group $SO(10) \times U(1)_\psi$. 
In addition to the $SO(10)$ ${\bf 16}$-plet SM fermions with a  $U(1)_\psi$ charge of $+1$, 
the model includes three generations of $SO(10)$ ${\bf 10}$-plets and $SO(10)$ singlet fermions with $U(1)_\psi$ charges $-2$ and $+4$, respectively. 
Each generation of these fermions can be embedded into a ${\bf 27}$ representation of the $E_6$ group, 
and hence our model is free from all the gauge and mixed gauge-gravitational anomalies.  
As previously mentioned, we consider a two-step SSB of the $SO(10)$ gauge group to the SM gauge symmetry, with the PS gauge symmetry appearing at an intermediate scale. 
The $U(1)_\psi$ symmetry is also broken at the intermediate scale by the vacuum expectation value (VEV) of a $SO(10)$ singlet Higgs field. 
This field is identified with the inflaton field which drives the IPI inflation in our model, such that all monopoles associated with the GUT and the PS SSBs are adequately diluted. 
After inflation, the inflaton decays into the SM particles to reheat the universe. 
We show that a suitable parameter choice yields a reheating temperature smaller than the PS SSB scale but large enough to thermalize the RHNs for successful baryognesis via leptogenesis \cite{leptogenesis2}. 
The $SO(10)$ group has a center ${\bf Z}_4$ with a subgroup ${\bf Z}_2$. 
In our model, all the Higgs representations are ${\bf Z}_2$-even, hence the ${\bf Z}_2$ symmetry remains unbroken even after the SSB down to the SM \cite{SO10Z2}, 
and as a result the lightest mass eigenstate among electrically neutral components  in the new ${\bf 10}$-plet and singlet fermions serves as a dark matter (DM) in our universe 
 (for an axion DM scenario in the context of $SO(10)$ models, see, for example, Ref.~\cite{S010nonSUSY}). 
If the DM particle is mostly composed of a $SO(10)$ singlet fermion, it communicates with the SM particles mainly through the SM Higgs portal interactions. 
We identify the allowed parameter region for this Higgs-portal fermion DM scenario, which will be fully explored by the direct DM detection experiments in the near future. 
In addition to the discussion about the IPI scenario and the DM scenario, 
we consider other phenomenological constraints and theoretical consistencies, 
such as successful gauge coupling unification, the proton decay constraint, 
and the stability of the effective SM Higgs potential.   
We identify a model parameter space for which our GUT model is phenomenologically viable and theoretically consistent.

The rest of this paper is organized as follows. 
In the next section, we define our $SO(10) \times U(1)_\psi$ GUT model. 
In Sec.~\ref{sec:S010Inf}, we first give a brief review of the IPI scenario and then implement the IPI in our model. 
We conclude the section with an evaluation of the reheating temperature after inflation. 
In Sec.~\ref{sec:GCUandPD}, we examine gauge coupling unification in the presence of the new fermions and Higgs fields, and we investigate its consistency with the current lower bound on the proton lifetime. 
In Sec.~\ref{sec:DM}, we discuss the DM scenario in our model. 
We identify a parameter region to reproduce the observed DM relic density that is consistent with the current direct DM detection bound. 
In Sec.~\ref{sec:HiggsStab}, we examine the stability of  the effective SM Higgs potential and find a parameter region which can stabilize the SM Higgs potential up to the PS SSB scale. 
Our conclusion are summarized in Sec.~\ref{sec:conc}.

\section{$SO(10) \times U(1)_\psi$}
\label{sec:model}

\begin{table}[t]
\begin{center}
\begin{tabular}{ll|c|c|c}
\textbf{}                      & & $SO(10)$ & $U(1)_\psi$ & ${\bf Z}_4$  \\ \hline
\multicolumn{1}{l|}{\multirow{3}{*}{\textbf{Fermions}}} 
&$ { 16}_{SM}^{(i)} $  &  {\bf 16}   & + 1 & $\omega$ \\
\multicolumn{1}{l|}{}
&$ {10}_{E}^{(i)} $    & {\bf 10}   & -- 2 & $\omega^2$ \\
\multicolumn{1}{l|}{}                                   
&$ {1}_{E}^{(i)} $    &  {\bf 1} & + 4 & 1  \\ \hline
\multicolumn{1}{l|}{\multirow{6}{*}{\textbf{Scalars}}}  
&$ {10}_{H} $    &  {\bf 10}   & -- 2 & $\omega^{2}$ \\
\multicolumn{1}{l|}{}                                   
&$ {45}_{H} $    & {\bf 45}   &   + 4 & 1 \\
\multicolumn{1}{l|}{}                                   
&$ {126}_{H} $    &  {\bf 126}  & + 2 & $\omega^2$  \\
\multicolumn{1}{l|}{}                                   
&$ {210}_{H} $    &  {\bf 210}   & \;\;\;0 & 1 \\ \cline{2-5} 
\multicolumn{1}{l|}{}                                   
&$ {\Phi}_A $    &  {\bf 1}   & + 4 & 1 \\
\multicolumn{1}{l|}{}                                   
&$ {\Phi}_B $    &  {\bf 1}   & -- 8 & 1 \\
\hline
\end{tabular}
\end{center}
\caption{
Particle contents of the $SO(10) \times U(1)_\psi$ model. Here, $\omega= e^{i \pi/2} = i $. 
}
\label{tab:PC}
\end{table}

The particle content of the $SO(10) \times U(1)_\psi$  model is listed in Table~\ref{tab:PC}.   
The model includes three generation of fermions in ${\bf 16}$ ($+1$), ${\bf 10}$ ($-2$), and ${\bf 1}$ ($+4$) representations of $SO(10)\times U(1)_\psi$. 
Each ${\bf 16}$-plet fermion (${16}_{\rm SM}^{(i)}$, $i = 1,2,3$) includes the $i$-th generation SM fermions plus one SM singlet RHN. 
The ${\bf 10}$-plets (${10}_{\rm E}^{(i)}$) and singlets (${1}_{\rm E}^{(i)}$) are new fermions. 
With the $U(1)_\psi$ charge assignments for the fermions in Table~\ref{tab:PC}, 
each generation of these fermions can be embedded into a ${\bf 27}$ representation of the $E_6$ group, 
and hence the model is free from all the gauge and mixed gauge-gravitational anomalies. 
Various representations of Higgs (scalar) fields are introduced in the table which break the $SO(10) \times U(1)_\psi$ group into the SM gauge group via the intermediate PS gauge group. 
The $SO(10)$ group has a center ${\bf Z}_4$, under which a ${\bf 16}$-plet transforms as ${\bf 16} \to i {\bf 16}$.
The ${\bf Z}_4$ charges of all other representations are fixed by this transformation law, 
which are listed in the last column of Table~\ref{tab:PC}. 
By VEVs of the various Higgs fields in the table, the ${\bf Z}_4$ symmetry is broken to its sub-group ${\bf Z}_2$ \cite{SO10Z2}. 
Under this ${\bf Z}_2$ symmetry all the particles except the SM ${16}_{SM}^{(i)}$ are ${\bf Z}_2$-even. 
Because of the ${\bf Z}_2$ symmetry, the lightest mass eigenstate among the ${\bf 10}$-plets and the singlet fermions 
is stable and hence a DM candidate. 
In fact, the DM candidate is stable even when higher dimensional operator are introduced because of the $SO(10)$ and Lorentz invariance (see, for example, Ref.~\cite{Ferrari:2018rey} for a variety of DM candidates in the $SO(10)$ scenario).

We assume a suitable Higgs potential for the Higgs fields listed in Table~\ref{tab:PC} such that their VEVs break $SO(10) \times U(1)_\psi$ to the SM gauge group. 
Consider the decomposition of Higgs representations under the PS gauge group of 
    $SU(4)_c \times SU(2)_L\times SU(2)_R$:      
\bea
{\bf 210} &=& ({\bf 1},{\bf 1},{\bf 1}) \oplus ({\bf 15},{\bf 1},{\bf 1}) \oplus ({\bf 6}, {\bf 2},{\bf 2}) \oplus ({\bf 15},{\bf 3},{\bf 1}) \oplus ({\bf 15},{\bf 1},{\bf 3}) \oplus ({\bf 10},{\bf 2},{\bf 2}) \oplus (\overline{\bf 10},{\bf 2},{\bf 2}) , \nonumber \\
{\overline {\bf 126}} &=& ({\bf 6},{\bf 1},{\bf 1}) \oplus ({\bf 10},{\bf 1},{\bf 3}) \oplus ({\overline {\bf 10}}, {\bf 3}, {\bf 1}) \oplus ({\bf 15}, {\bf 2}, {\bf 2}), \nonumber \\
{\bf 45} &=& ({\bf 1},{\bf 1},{\bf 3}) \oplus ({\bf 1},{\bf 3},{\bf 1}) \oplus ({\bf 6},{\bf 2},{\bf 2})\oplus ({\bf 15},{\bf1},{\bf1}), \nonumber \\
{\bf 10}&=& ({\bf 1},{\bf 2},{\bf 2}) \oplus ({\bf 6},{\bf 1},{\bf 1}). 
\label{eq:PSdecomp}
\eea 
We consider the following path for the SSBs:
\bea
SO(10) \times U(1)_\psi 
& \xrightarrow{\langle{210}_{H}\rangle}&
SU(4)_c \times SU(2)_L \times SU(2)_R \times U(1)_\psi \nonumber \\
& \xrightarrow[]{\langle{\overline {126}}_{H}\rangle,  \langle{45}_{H}\rangle, \; \langle{ \Phi}_{A,B}\rangle }& 
SU(3)_c \times SU(2)_L \times U(1)_{Y}  \nonumber \\
& \xrightarrow {\langle{10}_{H}\rangle}& 
SU(3)_c \times U(1)_{EM}. 
\label{eq:SB}
\eea
Here, the PS (and $U(1)_\chi$) singlet component of ${ 210}_{H}$, $({\bf 1},{\bf 1},{\bf 1})$, 
develops a GUT scale VEV ($ \langle {210}_{H} \rangle = M_{\rm GUT}$), 
which spontaneously breaks the $SO(10)$ gauge symmetry to the intermediate PS gauge group at the GUT scale. 
The PS gauge group is then spontaneously broken to the SM gauge group when $({\bf 10},{\bf 1},{\bf 3})$  of ${\overline {126}}_{H}$, $({\bf 15},{\bf1},{\bf1})$ of ${45}_{H}$ and ${\Phi}_{A,B}$ develop VEVs. 
For simplicity, we fix a common intermediate scale VEV for 
$\langle{\overline {126}}_{H}\rangle = \langle {45}_{H} \rangle = \langle{\Phi}_{A,B}\rangle  \equiv M_I$. 
Under the PS group decomposition, we assume that only the Higgs components listed in Table~\ref{tab:HiggsVEVs} have intermediate-scale masses while the other components have GUT-scale mass. 
The mass spectrum of the scalars ${\Phi}_{A,B}$ will be discussed later. 
Under the SM gauge groups, there are four Higgs doublets: 
    two in $({\bf 1},{\bf 2},{\bf 2})$ of ${10}_{H}$ and the  other two in $({\bf 15}, {\bf 2}, {\bf 2})$ of ${\overline {126}}_{H}$. 
We assume that all of these four Higgs doublets develop non-zero VEV at the electroweak scale, and only one linear combination of the doublets is light (doublet-doublet Higgs mass splitting) \cite{S010nonSUSY}. 
The light Higgs doublet is identified with the SM Higgs doublet, 
and the other linear combinations are heavy with masses of order $M_I$.\footnote{
The electroweak scale VEV for the $({\bf 15}, {\bf 2}, {\bf 2})$ can be realized by an induced VEV mechanism from a mixed scalar coupling ${126} \; {\overline {126}}\;  {126} \; {10}_H$ \cite{monopole1}.
} 
Following the $U(1)_\psi$ SSB, the $U(1)_\psi$ gauge boson ($Z^\prime$) acquires its mass which is given by
\bea 
m_{Z^\prime} \simeq \; g \; \sqrt{16 \langle { \Phi}_{A}\rangle^2 + 64 \langle {\Phi}_{B}\rangle^2 + 4 \langle \overline{{126}}_H \rangle^2+ 16 \langle { 45}_H \rangle^2} =  10 g {M_I},  
\label{eq:masses}
\eea
where $g$ is the $U(1)_\psi$ gauge coupling, $\langle {\Phi}_{A,B}\rangle = \langle{\overline {126}}_{H}\rangle =  \langle {45}_{H} \rangle = M_I$, and we neglect the contributions from the electroweak scale VEVs.

\begin{table}[t]
\begin{center}
\begin{tabular}{  c  |  c   }
&      $M_{I}$   
\\ \hline
$\overline{{126}}_H$  & $({\bf10},{\bf1},{\bf3})$,  $({\bf15},{\bf2},{\bf2})$
\\
${45}_H$  &  $({\bf15},{\bf1},{\bf1})$
\\
${10}_H$  &  $({\bf 1},{\bf 2},{\bf 2})$
\\ 
\end{tabular}
\end{center}
\caption{The Higgs mass spectrum; all other components have GUT scale masses.}
\label{tab:HiggsVEVs}
\end{table}

Let us consider fermion masses in our model. 
The Yukawa couplings for the SM fermions are given by 
\bea
 {\cal L} \supset   {16}_{SM} \left(Y_{10}  {10}_{H} + Y_{\overline{126}} {\overline {126}}_{H}  \right)  {16}_{SM}, 
\label{eq:SMY}
\eea
where the generation index has been suppressed. 
This is the so-called minimal $SO(10)$ model to generate realistic SM fermion mass matrices. 
Fitting of the fermion masses and flavor mixings is beyond the scope of the present work. 
We refer to Ref.~\cite{S010nonSUSY} for a detailed analysis of realistic fermion mass matrices. 
In Eq.~(\ref{eq:SMY}), the $U(1)_\psi$ gauge symmetry forbids Yukawa interaction of the form, ${16}_{SM} {10}_{H}^* {16}_{SM}$, which is generally allowed in non-SUSY $SO(10)$ models.
The Yukawa couplings of new fermions are given by 
\bea
 {\cal L} = &&\sum_i \frac{1}{2}Y_{A}^{(i)}{\Phi}_{A}  {10}_{E}^{(i)} {10}_{E}^{(i)} 
+ \sum_{i \neq j} \frac{1}{2}Y_{45}^{(ij)} {45}_H  {10}_{E}^{(i)} {10}_{E}^{(j)} 
  \nonumber \\
 &&+\sum_i \frac{1}{2} Y_{B}^{(i)} {\Phi}_{B}  { 1}_{E}^{(i)} { 1}_{E}^{(i)} 
+ \sum_{i, j} { Y_H}^{(ij)} {1}_{E}^{(i)}  { 10}_{E}^{(j)} {10}_{H}, 
\label{eq:ExoticY}
\eea 
where $Y_{45}^{(ij)}$ is anti-symmetric.  
The mass spectrum of the new fermions will be discussed in Sec.~\ref{sec:DM}.

\section{Inflation Scenario in $SO(10) \times U(1)_\psi$}
\label{sec:S010Inf}
As discussed in Sec.~\ref{sec:Intro}, a low-scale inflationary scenario with $H_{inf} < M_I$ is necessary to dilute the monopoles generated through the PS SSB at intermediate scale ($M_I$). 
In this section, we implement the IPI scenario in Ref.~\cite{IPI} to $SO(10) \times U(1)_\psi$ model and identify the parameter space to realize $H_{\rm inf} < M_{I}$. 
The $U(1)_\psi$ gauge symmetry is crucial for a successful IPI scenario, where the $SO(10)$ singlet Higgs field $\Phi_A$ is identified with the inflaton.

\subsection{Inflection-point Inflation }
For the reader's convenience, this sub-section is devoted to outline the general setup of the IPI scenario. 
See Ref.~\cite{IPI} for more details. 
The IPI is a low-scale inflation scenario driven by a single scalar field, in which the inflation potential exhibits an approximate inflection-point at a scale $M$.  
Consider the Taylor series of the inflaton potential $V (\phi)$ around $\phi=M$ up to the cubic term: 
\bea
V(\phi)\simeq V_0 +V_1 (\phi-M)+\frac{V_2}{2} (\phi-M)^2+\frac{V_3}{6} (\phi-M)^3, 
\label{eq:PExp}
\eea
   where $V_0 = V(M)$, and $V_n \equiv  {\rm d}^{n}V/{\rm d} \phi^n |_{\phi =M}$.  
It will soon be clear that higher order terms in the expansion can be neglected.

Using the potential of Eq.~(\ref{eq:PExp}), the inflationary slow-roll parameters at the scale $M$ are expressed as
\bea
\epsilon \simeq \frac{M_{P}^2}{2} \left( \frac{V_1}{V_0} \right)^2, \;\;
\eta \simeq M_{P}^2 \left( \frac{V_2}{V_0} \right), \;\;
\zeta^2 = M_{P}^4  \frac{V_1 V_3}{V_0^2}, 
\label{eq:IPa}
\eea
where $M_{P} = m_P/\sqrt{8\pi} = 2.43\times 10^{18}$ GeV is the reduced Planck mass. 
The inflationary predictions for the spectral-index ($n_s$), the tensor-to-scalar ratio ($r$), and the running of the spectral index ($\alpha$) are expressed in terms of the slow-roll parameters as 
\bea
n_s = 1-6\epsilon+2\eta, \; \; 
r = 16 \epsilon , \;\;
\alpha = 16 \epsilon \eta -24 \epsilon^2-2 \zeta^2.
 \label{eq:IPred}
\eea 
The amplitude of the scalar perturbation ($\Delta^2_{\mathcal{R}}$) is given by   
\begin{equation} 
\Delta_{\mathcal{R}}^2 = \frac{1}{24 \pi^2} \frac{V_0}{M_P^4 \epsilon }. 
 \label{eq:PSpec}
\end{equation}
Using the central values, $\Delta_{\mathcal{R}}^2= 2.195 \times 10^{-9}$ and $n_s = 0.9649$, from the Planck 2018 results \cite{Planck2018}, 
    we can express $V_1$ and $V_2$ as 
\bea
\frac{V_1}{M^3}&\simeq& 1.96 \times 10^3 \left(\frac{M}{M_P}\right)^3\left(\frac{V_0}{M^4}\right)^{3/2}, \nonumber \\
\frac{V_2}{M^2}&\simeq& -1.76 \times 10^{-2}  \left(\frac{M}{M_P} \right)^2 \left(\frac{V_0}{M^4}\right), 
\label{eq:FEq-V1V2}
\eea
where we have used $\epsilon(M) \ll |\eta(M)|$ in the IPI scenario \cite{IPI} in deriving the second equation. 

Slow-roll inflation takes place as the inflaton field slowly rolls down the inflaton potential from $\phi = M $ to $\phi = \phi_E < M$, 
where $\phi_E$ is the inflaton value at the end of inflation which is determined by $\epsilon (\phi_E) =1$.
As derived in Ref.~\cite{IPI}, the number of e-folding during inflation is approximately given by
\bea
N\simeq \pi \frac{V_0}{M_{P}^2\sqrt{2 V_1 V_3}} \; . 
\label{eq:CV4} 
\eea
To solve the horizon problem, we may set  $N = 50-60$. 
Using Eqs.~(\ref{eq:FEq-V1V2}) and (\ref{eq:CV4}), we express $V_3$ as  
\bea
\frac{V_3}{M} \simeq 6.99 \times 10^{-7} \; \left( \frac{60}{N} \right)^2  
   \left( \frac{M}{M_P} \right) \left( \frac{V_0}{M^4}   \right)^{1/2} . 
\label{eq:FEq-V3} 
\eea 
With the above expressions for $V_{1,2,3}$ in terms of $V_0$ and $M$, 
  we find the IPI predictions for $r$ and $\alpha$ as follows:  
\bea 
  r &=& 3.08 \times 10^7 \; \left( \frac{V_0}{M_P^4} \right), \nonumber \\
\alpha &\simeq& - 2\zeta^2 = -  2.74 \times 10^{-3}\left(\frac{60}{N}\right)^2.  
\label{eq:FEq-r}
\eea 
In the IPI scenario, the prediction for $\alpha$ is uniquely determined if $N$ is specified. 
For $N=60$, this prediction is consistent with $\alpha = - 0.0045\pm 0.0067$ from the Planck 2018 results \cite{Planck2018}. 
Precision measurement of $\alpha$ in future experiments can reduce the error to $\pm 0.002$ \cite{RunningSpectral}, 
   so that the IPI prediction can be tested in the foreseeable future.

\subsection{Inflection-point Inflation in $SO(10) \times U(1)_\psi$}
\label{sec:S010IPI}
Let us now implement the IPI scenario in the $SO(10) \times U(1)_\psi$ model 
   by identifying the $SO(10)$ singlet Higgs field $\Phi_{A}$ with the inflaton. 
Assuming $\Phi_A$ is very weakly coupled to the other Higgs fields, we consider the tree-level inflaton/Higgs potential given by 
\bea
  V_{tree} = \lambda \left( \Phi_A^\dagger \Phi_A - \frac{M_I}{2}  \right)^2 \simeq \frac{1}{4}\lambda \varphi^4,    
\eea
where $\varphi =\sqrt{2} \Re[ \Phi_A]$ is the real component of $\Phi_A$, and we identify it with the inflaton. 
To obtain the final expression for the inflaton potential, we have used $\varphi \gg M_I$ during inflation.

Taking quantum corrections into account, we consider a  renormalization group (RG) improved effective potential given by 
\bea
V(\varphi) = \frac{1}{4} \lambda (\varphi)\;\varphi^4. 
\label{eq:VEff}
\eea
Here, $\lambda (\varphi)$ is the solution to the following RG equations:
\bea
\varphi  \frac{d g}{d \varphi} &=& \frac{1}{16 \pi^2} \left( \frac{1448}{3} \right) g^3,         \nonumber\\
\varphi \frac{d Y_{A}^{(i)}}{d \varphi}   &=& \frac{1}{16 \pi^2}\left( 24 g^2 Y_{A}^{(i)} + Y_{A}^{(i)}\left(-48g^2  + \sum_j {Y_{A}^{(j)}}^2\right)\right),   
\nonumber\\
\varphi \frac{d \lambda}{d \varphi}  &=& \beta_{\lambda},
  \label{eq:RGEs}
\eea
where $ Y_{A}$s are the ${\bf 10}$-plet fermion Yukawa couplings, and the beta-function of the inflaton quartic coupling ($\beta_{\lambda}$) is given by
\bea
\beta_{\lambda} = \frac{1}{16 \pi^2} \left( 5 \lambda^2 + 
2 \lambda \left(- 48 g^2 + {Y_{A}^{(i)}}^2 + \sum_i {Y_{A}^{(i)}}^2 \right)  + 6144 g^4 - 4 \sum_i {Y_{A}^{(i)}}^4
 \right). 
\label{eq:BGen}
\eea
For simplicity, we have neglected the contribution of  $Y_{45}^{(ij)}$ to $\beta_{\lambda}$ by assuming sufficiently small $Y_{45}^{(ij)}$.

The constants $V_{1,2,3}$ in Eq.~(\ref{eq:PExp}) can be expressed in terms of $\lambda$ and $\beta_\lambda$ as follows: 
\bea
\frac{V_1}{M^3}&=& \left.\frac{1}{4} (4 \lambda + \beta_\lambda)\right|_{\varphi= M},\nonumber \\
\frac{V_2}{M^2}&=&  \left.\frac{1}{4} (12\lambda + 7\beta_\lambda+M \beta_\lambda^\prime)\right|_{\varphi= M}, \nonumber \\
\frac{V_3}{M}&=&  \left.\frac{1}{4} (24\lambda + 26\beta_\lambda+10M \beta_\lambda^\prime+M^2 \beta_\lambda^{\prime\prime})\right|_{\varphi= M}, 
\label{eq:ICons2}
\eea
where the prime denotes $d/d\varphi$.
In order for the effective inflation potential to exhibit an approximate inflection-point at $M$, 
we require $V_1/M^3\simeq 0$ and $V_2/M^2\simeq 0$, so that 
\bea
 \beta_\lambda (M)\simeq -4\lambda(M), \qquad
 M\beta_\lambda^{\prime}(M)\simeq -12 \lambda(M) - 7\beta_\lambda(M) \simeq  
 16 \lambda (M). 
 \label{eq:Cond1}
\eea
For $g (M) , Y_A^{(i)} (M), \lambda (M)< 1$, we can approximate $M^2 \beta_\lambda^{\prime\prime}(M)  = - M \beta_\lambda^{\prime}(M) + (M \beta_\lambda^{\prime}(M))^2 \simeq - M \beta_\lambda^{\prime}(M)$, where we have neglected contributions from higher order terms, namely ${\cal O}(g^8)$, ${\cal O}((Y_A^{(i)})^8)$ and ${\cal O}(\lambda^4)$. 
Together with the relations in Eq.~(\ref{eq:Cond1}), it simplifies the last equation in Eq.~(\ref{eq:ICons2}) to $V_3/M \simeq 16 \;\lambda(M)$. 
Using Eq.~(\ref{eq:FEq-V3}) and $V_0\simeq (1/4) \lambda(M) M^4$, 
   we arrive at 
\bea
\lambda(M)\simeq 4.8 \times 10^{-16} \left(\frac{M}{M_{P}}\right)^2\left(\frac{60}{N}\right)^4. 
\label{eq:FEq1} 
\eea 
For the rest of the analysis, we set $N=60$. 
With the inflaton quartic coupling determined by $M$, we express the tensor-to-scalar ratio ($r$) and the Hubble parameter during the inflation ($H_{inf}$) as 
\bea
r &\simeq& 3.7 \times 10^{-9}  \left(\frac{M}{M_{P}}\right)^6, 
\nonumber \\
H_{inf} &\simeq& \sqrt{\frac{V_0}{3 {M_P}^2}}\simeq 1.5\times 10^{10} \;{\rm GeV} \;\left(\frac{M}{M_P}\right)^3.  
\label{eq:FEqR} 
\eea 
Note that $H_{inf} \lesssim 10^{10}$ GeV for $M \lesssim M_P$. 
In Ref.~\cite{IPI}, an upper-bound $M \lesssim 5.7 M_P$ has been obtained from theoretical consistency. 
In the following sections, we will find $M_I=10^{11}-10^{12}$ GeV in our model. 
Therefore, the monopole problem is solved by taking $M \lesssim M_P$. 
For $M \lesssim M_P$, the prediction of the tensor-to-scalar ratio $r < 3.7 \times 10^{-9}$ is much smaller than the current upper bound of $r\ll  0.065$ from the Planck 2018 observation \cite{Planck2018}.

The conditions in Eq.~(\ref{eq:Cond1}) to realize the (approximate) inflection-point at $M$ allows us to derive a relation between the gauge and Yukawa couplings. 
For simplicity, we assume $Y_{A}^{(1,2)} \ll Y_{A}^{(3)} \equiv Y$. 
Since the gauge and Yukawa couplings are independent of $\lambda$, we also assume $g,  Y_{A}^{(3)} \gg \lambda$.
In this case, the first condition in  Eq.~(\ref{eq:Cond1}) with the very small $\lambda$ in Eq.~(\ref{eq:FEq1}) leads to $\beta_\lambda(M) \simeq 0$ such that  
\bea
Y(M)\simeq 6.3\;g(M). 
\label{eq:FEq3}
\eea
Employing this relation and explicitly evaluating the second condition in Eq.~(\ref{eq:Cond1}) using the RG equations in Eq.~(\ref{eq:RGEs}), we find a relation, $\lambda(M)\simeq 26 \, g(M)^6$.  
Thus, we can express the $U(1)_\psi$ gauge coupling as 
\bea
g(M)\simeq 1.6\times 10^{-3} \;\left(\frac{M}{M_{P}}\right)^{1/3}.
\label{eq:FEq2} 
\eea
Thus, all couplings at the scale $M$, namely $g(M)$, $Y(M)$ and $\lambda(M)$ are determined in terms of $M/M_P$.

\begin{figure}[t]
\begin{center}
\includegraphics[scale=0.88]{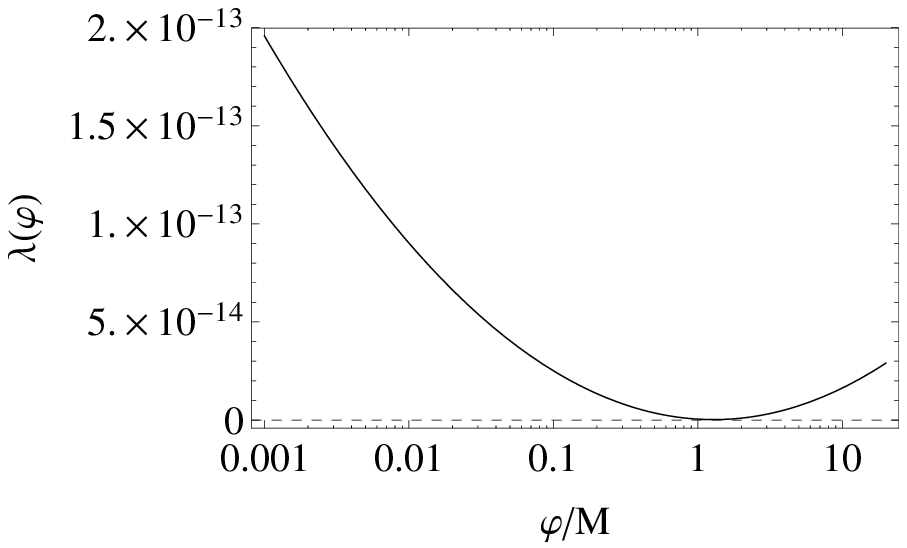} \;
\includegraphics[scale=0.62]{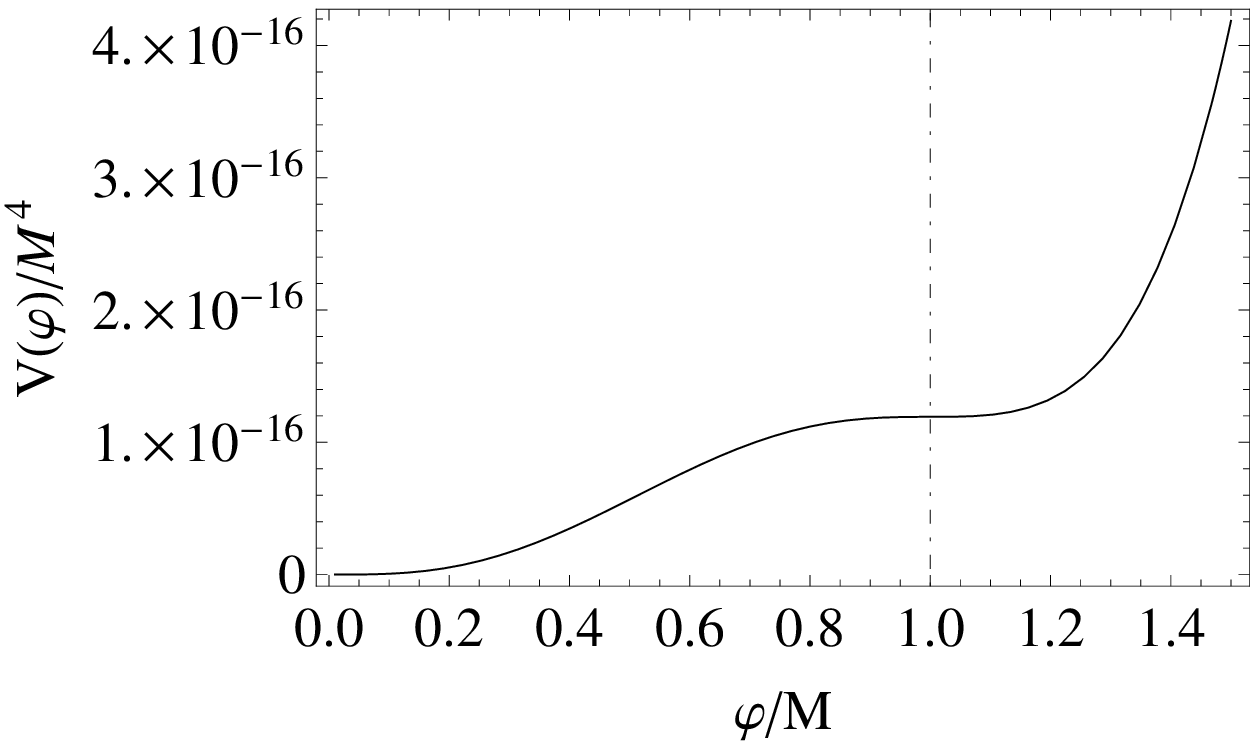}       
\end{center}
\caption{
The left panel shows the RG running of inflaton quartic coupling as a function of $\varphi/M$.
We have fixed $M = M_{P} $, so that  $g (M)=1.6 \times 10^{-3}$, $Y (M) \simeq 1.0 \times 10^{-2}$, and $\lambda(M) \simeq 4.8 \times10^{-16}$. 
Dashed horizontal line corresponds to $\lambda=0$. 
The right panel shows the RG improved effective inflaton potential with an (approximate) inflection-point at $\varphi \simeq M$. 
}
\label{fig:InfPot}
\end{figure}

Next we evaluate the low-energy values of  $g(\varphi)$,  $Y(\varphi)$ and $\lambda (\varphi)$ 
   by solving the RG equations. 
Because of $g(M), Y(M) \ll1$, it is easy to find the approximate solutions to their RG equations: 
\bea
g(\varphi)   &\simeq& g(M) +\beta_{g}(M) \ln \left[\frac{\varphi}{M}\right]  ,\nonumber \\ 
Y(\varphi)  &\simeq& Y(M)+ \beta_{Y}(M) \ln \left[\frac{\varphi}{M}\right], 
\label{eq:gYatVEV}
\eea
where $\beta_{g}(M)$ and $\beta_{Y}(M)$ are the beta-functions of $g$ and $Y$ evaluated at $M$, respectively. 
Since $\lambda (M)$ is extremely small,  
  $\beta_{\lambda}$ is mainly controlled by the gauge and the Yukawa couplings, 
\bea
 \beta_{\lambda}(\varphi) 
&\simeq& \frac{1}{16 \pi^2}\left( 6144\; g(\varphi) ^4 - 4\; Y(\varphi) ^4\right) \nonumber \\
&\simeq&\frac{1}{16 \pi^2} \times 16 \lambda (M) \ln \left[\frac{\varphi}{M}\right], 
\label{eq:betaL}
\eea
where we have used $\beta_\lambda (M) = 0$ and Eq.~(\ref{eq:gYatVEV}). 
Hence, we find the approximate solution, 
\bea
\lambda(\varphi) \simeq 3.8\times 10^{-15}\left(\frac{M}{M_P}\right)^2\left(\ln \left[\frac{\varphi}{M}\right] \right)^2. 
\label{eq:Lmu} 
\eea
At the $U(1)_\psi$ symmetry breaking scale $\varphi= M_I$, 
  we obtain the mass spectrum: 
\bea
m_{Z^\prime} &\simeq&  10 g(M_I) M_I \simeq  2.3 \times 10^{-2} \times M_I \;\left(\frac{M}{M_{P}}\right)^{1/3}, 
\nonumber \\
m_{\varphi}&=& \sqrt{ 2 \lambda(M_I)} M_I \simeq  8.7\times 10^{-8} \times  M_I \left|\ln \left[\frac{M}{M_I}\right]\right| \left(\frac{M_I}{M_P}\right), 
\nonumber \\
m_{10}^{(3)} &\simeq& \frac{1}{2}Y(M_I) M_I \simeq \frac{1}{3} m_{Z^\prime}. 
\label{eq:mA}
\eea 
In the following analysis, we fix $M = M_P$ for simplicity, so that the mass spectrum is uniquely determined by $M_I$.

In Fig.~\ref{fig:InfPot}, we plot the running quartic coupling (left) and the RG improved effective inflaton potential (right). 
Here, $\lambda(M)\simeq 4.8 \times10^{-16}$, $g(M) \simeq 1.6\times 10^{-3}$, and $Y(M)\simeq  1.0 \times 10^{-2}$ 
  with our choice of $M = M_P$. 
In the left panel, the running quartic coupling shows a minimum at $\varphi \simeq M$. 	
In the right panel, we can see that the inflaton potential exhibits an (approximate) inflection-point 
  at $\varphi \simeq M$ (marked as the vertical dashed-dotted line).

\subsection{Reheating Temperature and Thermal Leptogenesis}
\label{sec:Reheat}
To connect our inflation scenario with the Standard Big Bang cosmology, we consider reheating after inflation. 
After the end of inflation, the inflaton rolls down to the potential minimum and then oscillates around the minimum. 
As the age of the universe reaches the lifetime of the inflaton, the latter decays to the SM particles and the total inflaton energy is transmitted to SM particles as radiation.  
Assuming that the decay products are instantly thermalized, 
   we estimate the reheat temperature ($T_R$) by  
\begin{eqnarray}
    T_R \simeq \left(\frac{90}{\pi^2 g_*}\right)^{1/4} \sqrt{\Gamma M_P},  
\label{eq:TR}
\end{eqnarray} 
where $\Gamma$ is the decay width of the inflaton and $g_*$ is the total number of degrees of freedom of the thermal plasma. 
We may express the decay width of inflaton as  
\bea
\Gamma  \simeq 1.4 \; {\rm GeV} \times \sqrt{g_*} \left(\frac{T_R [{\rm GeV}]}{10^{10}}\right)^2 .  
\label{eq:gamma1}
\eea

For a coupling between the inflaton and the SM particles, 
  we consider the following gauge invariant coupling in the scalar potential: 
\bea
V\supset \Lambda \Phi_A 10_H^2 \supset \Lambda \varphi  H_u H_d \supset \frac{\Lambda \sin2\beta}{2} \varphi H^\dagger H, 
\label{eq:InfPot}
\eea 
where $\Lambda > 0$ is a free mass parameter, 
   and $10_H \supset ({\bf 1},{\bf 2},{\bf 2}) = H_u \; ({\bf 1},{\bf 2}, +1/2)$ $\oplus H_d \; ({\bf 1},{\bf 2}^*, -1/2)$. 
The SM Higgs doublet (H) is realized as a linear combination of  $H_u$ and $H_d$, 
  and $H$ is embedded in $H_{u,d}$ as $H_{u} \supset  H \sin\beta$ and $H_{d} \supset { H}^\dagger \cos\beta$, 
   where $\tan\beta = v_u/v_d$ is a ratio of $H_u$ and $H_d$ VEVs. 
The decay width of the inflaton into a pair of SM Higgs doublets is given by
\bea
\Gamma (\varphi \to H^\dagger H ) \simeq  \frac{\Lambda^2 \sin^{2}2\beta}{4\pi \, m_\varphi}, 
\label{eq:gamma2}
\eea
where we have neglected the Higgs doublet mass. 
For $M = M_P$ and $M_I = 1.3 \times 10^{11}$ GeV, 
  we obtain $m_{\varphi} = 1.9\times 10^{5}$ GeV from Eq.~(\ref{eq:mA}), and thus the reheating temperature, 
\bea
T_R \simeq 10^{10} \;{\rm GeV} \left(\frac{\Lambda [{\rm GeV}]}{4.2 \times 10^5}\right), 
\label{Lambda}
\eea
with $g_* = 100$ and $\beta = \pi/3$. 
Although the inflaton can also decay into a pair of SM Higgs doublets also through the quartic coupling $ \lambda_{mix}\Phi_A^\dagger \Phi_A 10_H^\dagger 10_H$, we have assumed this small. 
Another possibility for the inflaton decay is through the Yukawa coupling in Eq.~(\ref{eq:ExoticY}) if a ${\bf 10}$-plet fermion is light enough. 
Since the infalton mass is much smaller than $M_I$, the Yukawa coupling $Y_A^{(i)}$ is very small 
   whenever a ${\bf 10}$-plet fermion is lighter than the inflaton.
Thus, we neglect the partial decay width of the inflaton for this process.

In our scenario, the Majorana RHN masses are generated by the PS SSB at the intermediate scale. 
This scale is a narural scale for the seesaw mechanism to generate light neutrino masses as well as thermal leptogenesis. 
As pointed out in Ref.~\cite{leptogenesis2}, 
there is a lower bound on the lightest RHN mass $\gtrsim 10^9$ GeV for a successful thermal leptogenesis scenario.
If the lightest RHN mass to be $10^{9}$ GeV, we may adjust $\Lambda=4.2 \times 10^5$  GeV in Eq.~(\ref{Lambda}) so that $10^9 < T_R[{\rm GeV}] \simeq 10^{10} < M_I$ for  successful thermal leptogenesis and also avoid a restoration of the PS symmetric vacuum.

\section{Gauge Coupling Unification and Proton Decay}
\label{sec:GCUandPD}
As discussed before, the $SO(10)$ breaking to the SM proceeds in two-steps. 
In the bottom-up picture, the SM gauge groups are first unified into the PS gauge group $SU(4){_c} \times SU(2)_L \times SU(2)_R$ at the intermediate scale $M_I$, 
and then the PS gauge group is unified into the $SO(10)$ group at $M_{GUT}$. 
In this section, we examine the RG evolutions of the gauge couplings 
and determine the mass spectrum of the new particles in order to realize the successful gauge coupling unification.  
We also consider a lower bound on $M_{GUT}$ from the current experimental lower bound 
   on the proton lifetime.

We consider the contribution of new particles to the RG running of the gauge couplings. 
For the Higgs sector, the fields listed in Table~\ref{tab:HiggsVEVs} contribute to the RG evolutions of the gauge couplings above the PS SSB scale, 
while only the SM Higgs doublet contributes to the RG equations of the SM gauge couplings below the PS SSB scale. 
The new $10_{E}$ fermion decomposition under the PS gauge group is given by Eq.~(\ref{eq:PSdecomp}). 
Under the SM gauge group, 
\bea
{10}^{(i)}_{E} =  D^{(i)} \; ({\bf1},{\bf2},+1/2) \oplus {\bar D}^{(i)} \; ({\bf1}, \bar{\bf2}, -1/2) \oplus T^{(i)} \; ({\bf3},{\bf1}, +1/3) \oplus {\bar T}^{(i)} \; ({\bar {\bf3}},{\bf1}, -1/3), 
\label{eq:10decomp}
\eea
where $D^{(i)}$ and ${\bar D}^{(i)}$ ($T$ and ${\bar T}^{(i)}$) are the SM $SU(2)_L$ doublets ($SU(3)_c$ triplets).   
In the previous section, we have fixed the ${10}^{(3)}_{E}$ fermion mass ($m_{10}^{(3)}$) in Eq.~(\ref{eq:mA}) 
   by the IPI analysis. 
For the other two ${\bf 10}$-plet fermions, we consider a mass splitting between the doublets and triplets 
   (the origin of the mass splitting will be discussed in the next section). 
It will turn out that this mass splitting is crucial to keep the unification scale $M_{GUT} < M_{P}$.

Let us now examine the RG evolution of the gauge couplings by solving their RG equations at the 1-loop level. 
For energy scale $\mu$ below the PS SSB scale ($\mu < M_I$), 
  the running SM gauge couplings obey the following RG equations: 
\bea
 \mu  \frac{d \alpha_{1}}{d \mu} &=& \frac{1}{2\pi}\alpha_{1}^2
\left(\frac{41}{10}+ \sum_{j= 1,2} \frac{2}{5} \theta (\mu - m_{D}^{(j)})+ \sum_{j= 1,2} \frac{4}{15}\theta (\mu - m_{T}^{(j)}) +  \frac{2}{3} \theta (\mu -m_{10}^{(3)})\right), 
 \nonumber \\
 \mu  \frac{d \alpha_{2}}{d \mu} &=& \frac{1}{2\pi}\alpha_{2}^2 \left(-\frac{19}{6} + \sum_{j= 1,2}\frac{2}{3} \theta (\mu - m_{D}^{(j)})+  \frac{2}{3} \theta (\mu - m_{10}^{(3)})\right), 
\nonumber \\
 \mu  \frac{d  \alpha_{3}}{d \mu} &=& \frac{1}{2\pi}\alpha_{3}^2 \left(-7+  \sum_{j= 1,2}\frac{2}{3}\theta (\mu - m_{T}^{(j)}) +  \frac{2}{3}\theta (\mu - m_{10}^{(3)}) \right). 
\label{eq:betafun1}   
\eea
Here, $\alpha_{2,3} = g_{2,3}^2/4\pi$ with $g_{2,3}$ being the $SU(3)_c$ and $SU(2)_L$ gauge couplings, respectively, 
  $\alpha_1 = g_1^2/4\pi$ with $g_{1} = \sqrt{5/3}\; g_Y$ and $g_Y$ the  $U(1)_Y$ gauge coupling, 
$\theta$ is a Heaviside function, $m_{10}^{(3)} \simeq 7.7 \times 10^{-3} M_I$ is fixed from Eq.~(\ref{eq:mA}) 
with $M=M_P$, 
and $m_{D}^{(j)}$  ($m_{T}^{(j)}$) are the doublet (triplet) component masses of the two ${\bf 10}$-plet fermions.
In the following analysis, 
we fix $m_{D}^{(1)}= m_{D}^{(2)} \equiv m_D$ and $m_{T}^{(1)}= m_{T}^{(2)} \equiv m_T$ ($m_{D, T} < M_I$), for simplicity.   
In solving the RG equations, we employ the SM gauge couplings at $\mu = m_t = 172.44$ GeV \cite{Buttazzo:2013uya}: 
\bea
g_{1} (m_t) = \sqrt{5/3} \times 0.35830, \qquad
g_{2} (m_t) =  0.64779 , \qquad 
g_{3} (m_t) = 1.1666.  
\eea 
In our analysis, $m_{D,T}$ and $M_I$ are free parameters.

For $M_I < \mu < M_{GUT}$, our theory is based on the PS gauge group. 
The relationship between the SM and the PS gauge couplings at $\mu = M_I$ are given by the tree-level matching conditions:   
\bea
 \alpha_{2} (M_I) = {\alpha}_{L} (M_I) ,
 \qquad \alpha_{3} (M_I) = {\alpha}_{4} (M_I) , 
 \qquad
 \alpha_{1}^{-1} (M_I) = \frac{3}{5} {\alpha}_{R}^{-1} (M_I) + \frac{2}{5}{\alpha}_{4}^{-1}(M_I), 
\eea
where $\alpha_{4,L,R}$ represent the gauge couplings of the gauge groups, $SU(4){_c}$, $SU(2)_L$, and $SU(2)_R$, respectively.  
With the initial values of the PS gauge couplings fixed by the matching conditions, 
  we solve the following RG equations of the PS gauge couplings for $M_I < \mu < M_{GUT}$: 
\bea
\mu  \frac{d {\alpha}_{4}}{d \mu} &=& \frac{1}{2\pi}{\alpha}_{4}^2 
\left(+1\right), 
\nonumber \\
 \mu  \frac{d {\alpha}_{L}}{d \mu} &=& \frac{1}{2\pi}{\alpha}_{L}^2 \left(4\right), 
\nonumber \\
 \mu  \frac{d {\alpha}_{R}}{d \mu} &=& \frac{1}{2\pi}{\alpha}_{R}^2
\left(\frac{32}{3}\right). 
\label{eq:betafun2}   
\eea
Here, the beta-functions include the contribution from all SM fermions, ${10}^{(i)}_{E}$ ($i= 1,2,3$) fermions, 
   the Higgs fields listed in Table~\ref{tab:HiggsVEVs}, and the PS gauge bosons.

\begin{figure}[t]
\begin{center}
\includegraphics[scale =0.8]{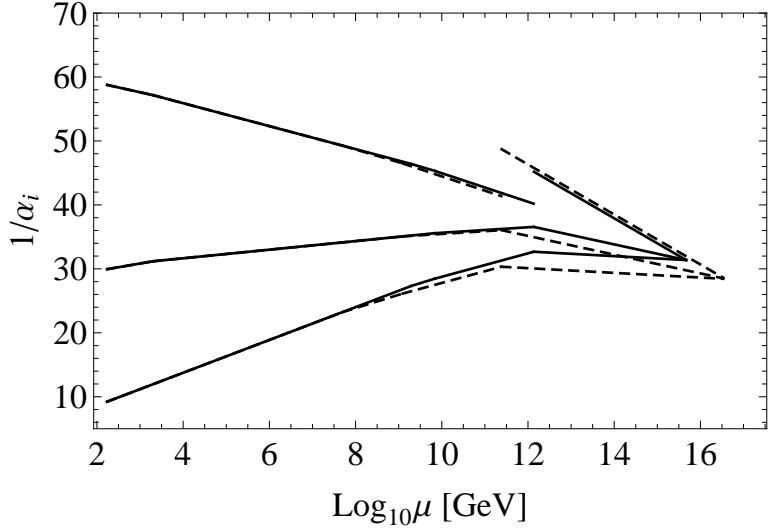} 
\end{center}
\caption{
For $m_{D} = 2$ TeV and $m_T =  5\times10^4$ TeV (solid lines) and $m_{T} = 2\times10^6$ TeV (dashed lines), 
the three solid and dashed lines from top to bottom correspond to $\alpha_{{1,2,3}}$ for $\mu < M_I$ and ${\alpha}_{{R,L,4}}$ for $M_I< \mu < M_{GUT}$. 
For $m_{T} = 5\times10^4 \; (2\times10^6)$ TeV, we find $M_I \simeq  1.4 \times 10^{12} \; (2.4 \times 10^{11})$ GeV and $M_{GUT} \simeq 4.6 \times 10^{15} \; (3.4 \times 10^{16})$ GeV. 
}
\label{fig:gcu}
\end{figure}

The analytic solutions for the above RG equations at scale $\mu$ are obtained as functions of three free parameters, $m_{D}$, $m_{T}$, and $M_I$.  
Next, we require gauge coupling unification at $\mu=M_{GUT}$: 
$ \alpha_{L} (M_{GUT}) =  \alpha_{R} (M_{GUT}) =  \alpha_{4} (M_{GUT}) \equiv \alpha_{GUT}$. 
With four free parameters,  $m_{D}$, $m_{T}$, $M_I$ and $M_{GUT}$, 
we can always find a solution to satisfy the gauge coupling unification condition.  
Once we fix the values of $m_{D}$ and $m_{T}$, 
the mass scales $M_I$ and $M_{GUT}$ are determined from the unification condition.   
In Fig.~\ref{fig:gcu}, we plot the RG running of the gauge couplings for a fixed value of $m_{D} = 2$ TeV and two different values of $m_T =  5\times10^4$ TeV (solid lines) and $m_{T} = 2\times10^6$ TeV (dashed lines).  
The three solid lines from top to bottom correspond to $\alpha_{{1,2,3}}$ for $\mu < M_I$ and ${\alpha}_{{R,L,4}}$ for $M_I< \mu < M_{GUT}$. 
For $m_{T} = 5\times10^4 \; (2\times10^6)$ TeV, we find $M_I \simeq  1.4 \times 10^{12} \; ( 2.4 \times 10^{11})$ GeV and $M_{GUT} \simeq 4.6 \times 10^{15} \; (3.4 \times 10^{16})$ GeV. 
The plot shows that as we increase the triplet fermion mass $m_T$, $\alpha_{GUT}$ and $M_{GUT}$ values decrease 
  while the $M_I$ value increases.

\begin{figure}[t]
\begin{center}
\includegraphics[scale =0.65]{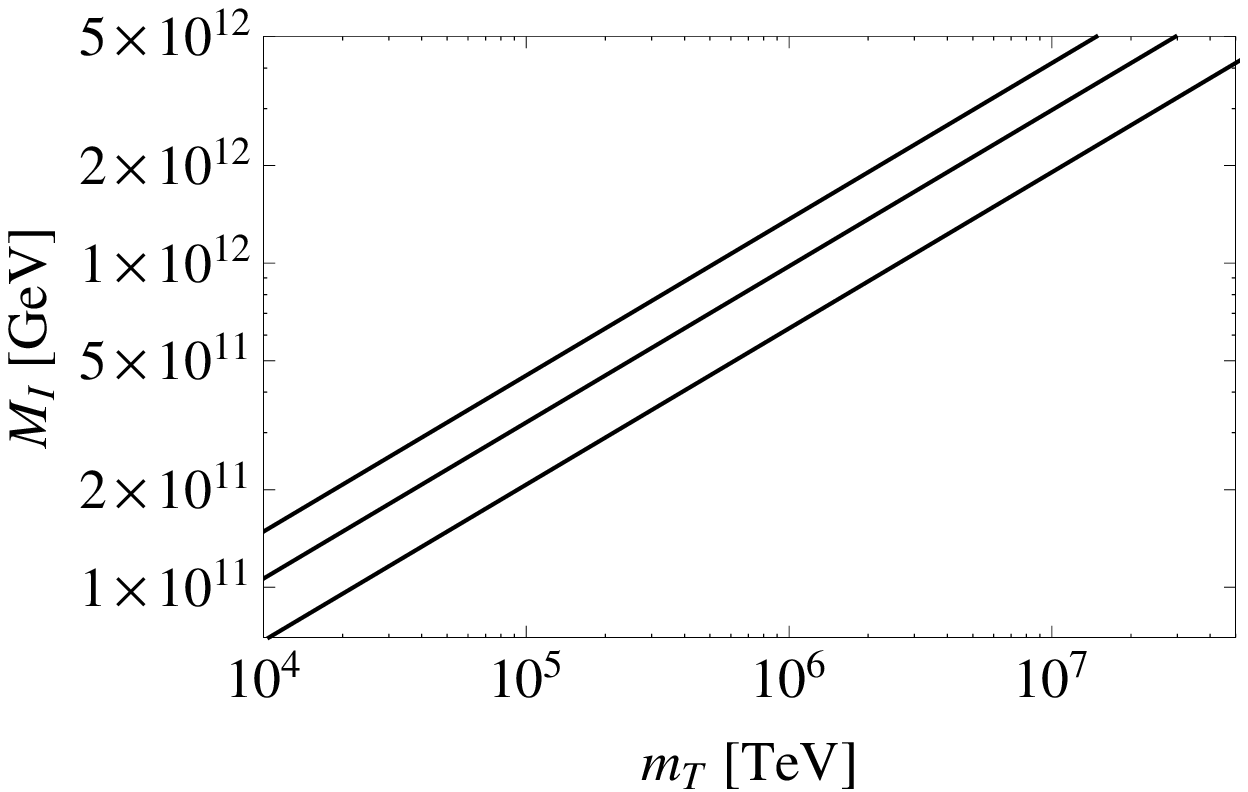} \;
\includegraphics[scale=0.58]{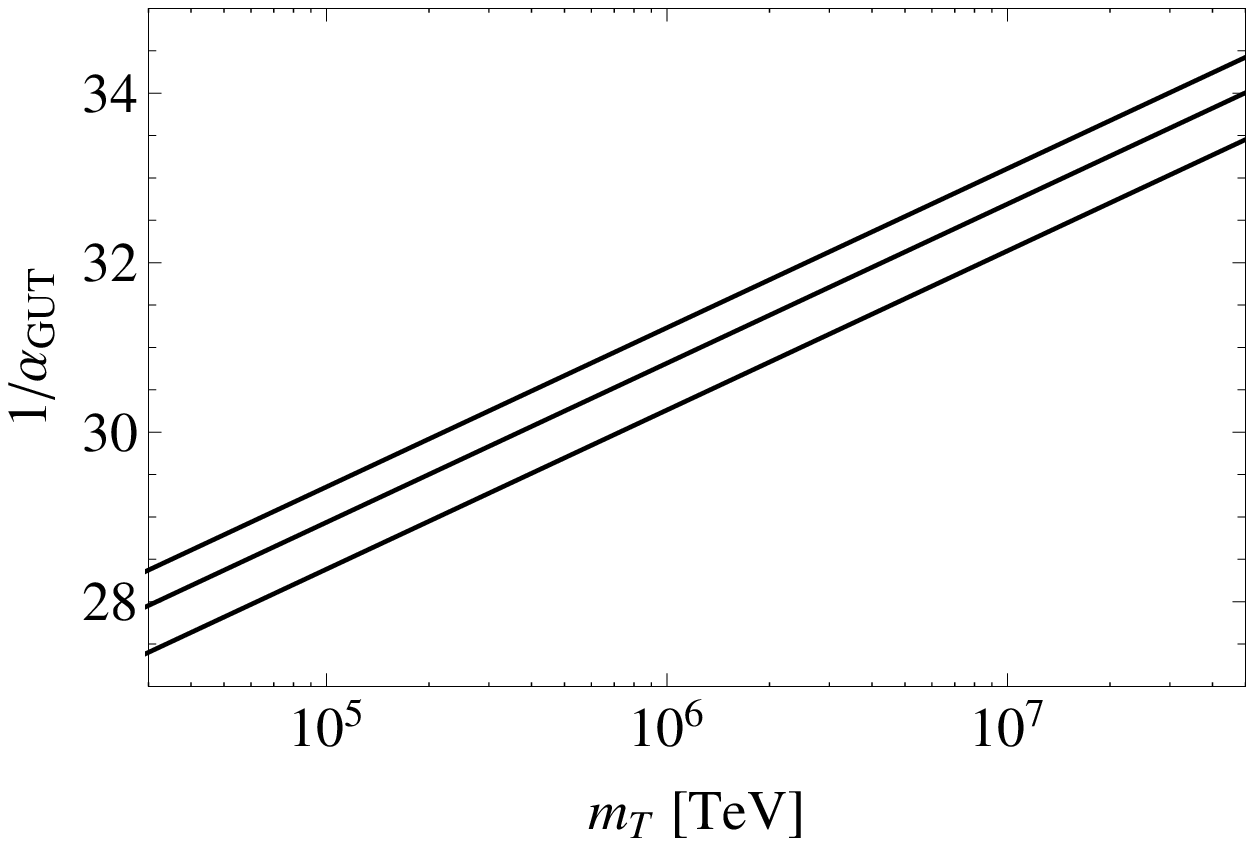}\\
\includegraphics[scale=0.95]{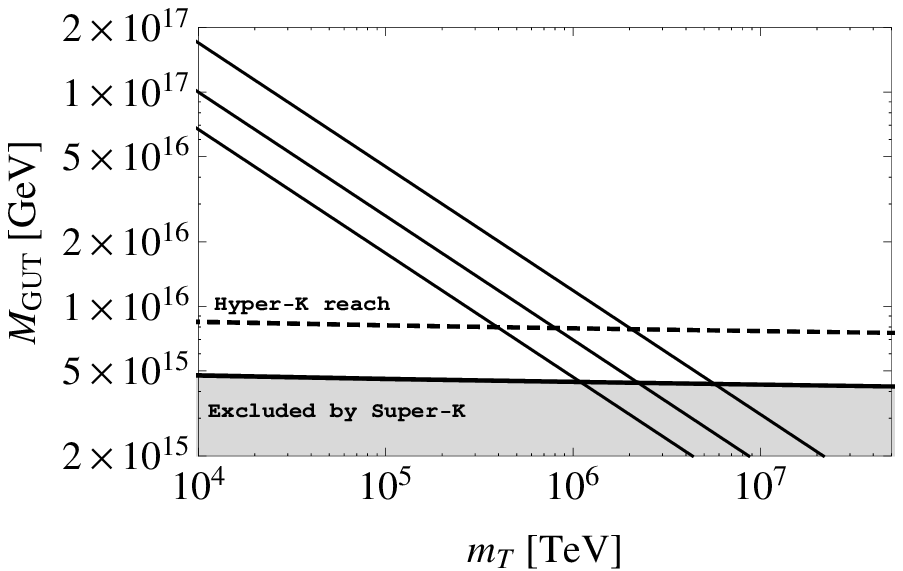}
\end{center}
\caption{
Top-left and top-right panels show $M_{I}$ and $1/\alpha_{GUT}$ as a function of $m_{T}$, respectively, 
for $m_{D} = 1$ TeV, $2$ TeV, and $5$ TeV (solid lines from top to bottom). 
The bottom panel shows $M_{GUT}$ as a function of $m_{T}$ for $m_{D} = 1$ TeV, $2$ TeV, and $5$ TeV (solid lines from top to bottom). 
The gray shaded region is excluded by the lower bound on proton lifetime from the Super-K result.  
The search reach of proton lifetime by the future Hyper-K experiment, $\tau_{HK} \simeq 10 \times \tau_{SK}$ \cite{Abe:2011ts}, is depicted as the dashed line. 
}
\label{fig:MIandMGUT}
\end{figure}

Since quarks and leptons are unified into a representation of the unified gauge group and baryon number is broken, 
proton decay is a typical prediction of GUTs.  
In our model, the main proton decay process, $p \to \pi^0 e^+$, 
is mediated by the $SO(10)$ GUT gauge bosons and the colored Higgs bosons in  $10_H$. 
For the GUT gauge boson mediated process, the proton lifetime is estimated 
  as $\tau_p \simeq (1/\alpha_{GUT}^2) M_{GUT}^4/m_p^5$ \cite{Nath:2006ut} 
  in terms of the unified gauge coupling $\alpha_{GUT}$, the gauge coupling unification scale $M_{GUT}$ 
  and the proton mass $m_p = 0.983$ GeV. 
For the colored Higgs mediated process, we estimate the proton lifetime 
  as $\tau_p \simeq m_{HC}^4/\left(m_p^5 Y^2_u Y^2_d \right)$  \cite{Nath:2006ut}, 
  where $Y_{u,d}  \simeq 10^{-5}$ are the up and down quark Yukawa couplings, and $m_{HC}$ 
  is a colored Higgs boson mass. 
Employing the lower bound on the proton lifetime for the process $p \to \pi^0 e^+$ 
    by the Super-Kamionkande (Super-K) experiment, $\tau_{SK}> 1.6 \times 10^{34}$ years \cite{Miura:2016krn}, 
    we find $M_{GUT}/\sqrt{\alpha_{GUT}} > 2.5 \times 10^{16}$ GeV 
    and $m_{HC} > 4.5 \times 10^{11}$ GeV 
    for the GUT gauge boson and the colored Higgs mediated processes, respectively. 
The proton decay bound constrains the parameter region for $m_{D}$ and $m_{T}$. 
We have taken $m_{HC} =M_{GUT}$ for the analysis in this section. 
However, our result for the gauge coupling unification remains almost the same 
    even for $4.5 \times 10^{11}$ GeV$ < m_{HC} <M_{GUT}$, 
    since the colored Higgs contribution to the beta-functions is not large.

In Fig.~\ref{fig:MIandMGUT}, we show our results for the gauge coupling unification 
   for various values of $m_D$ and $m_T$. 
The top panels depict  $M_{I}$ (left panel) and $1/\alpha_{GUT}$ (right panel) as a function of $m_{T}$ 
   for three fixed values of $m_{D} = 1$ TeV, $2$ TeV, and $5$ TeV from top to bottom. 
Gauge coupling unification is realized along the solid lines. 
In the bottom panel, we show $M_{GUT}$ as a function of $m_{T}$ 
   for $m_{D} = 1$ TeV, $2$ TeV, and $5$ TeV from bottom to top, respectively.  
The gray shaded region is excluded by the Super-K result. 
Note that the Super-K constraint leads to an upper bound on the triplet fermion mass, 
   $m_{T} < 2\times 10^6$ TeV, $8\times 10^5$ TeV and $4\times10^5$ TeV, respectively, 
   for  $m_{D} = 5$ TeV, $2$ TeV, and $1$ TeV. 
The search reach of the proton lifetime by the future Hyper-Kamiokande (Hyper-K) experiment, 
   $\tau_{HK} \simeq 10 \times \tau_{SK}$ \cite{Abe:2011ts}, is depicted as the dashed line.

We conclude this section with a comment on the result for the degenerate mass spectrum, $m_D^{(i)}= m_T^{(i)}$ ($i=1,2$). 
In this case, we find $M_I \simeq 1.7\times 10^9$ GeV and $M_{GUT}\simeq 1.4 \times 10^{19}$ GeV. 
This result is independent of the degenerate mass spectrum, 
since the ${\bf 10}$-plet fermions contribute to the gauge coupling beta-functions as complete $SO(10)$ multiplets. 
As shown in Figs.~\ref{fig:gcu} and \ref{fig:MIandMGUT},   the mass splitting lowers the gauge coupling unification scale 
starting from the Planck scale.

\section{Dark Matter in $SO(10) \times U(1)_\psi$}
\label{sec:DM}
Because of the residual ${\bf Z}_2$ symmetry after the SSB, 
the lightest mass eigenstate from a linear combination of the $SO(10)$ singlet fermions and the ${\bf 10}$-plet fermions is stable and is a suitable DM candidate if it is electrically and color neutral. 
In this section we consider the DM physics in our model. 
In the SM gauge group decomposition, the DM candidate is a linear combination of SM singlet and the $SU(2)_L$ doublet fermions, the so-called ``singlet-doublet DM''  (SD-DM) scenario \cite{SDDM}.
In the following, we identify the allowed parameter region to reproduce the observed DM relic density while satisfying the constraint from the direct DM detection experiments.

\subsection{Doublet-triplet Fermion Mass Splitting and Triplet Fermions Lifetime}
\label{sec:DTS}
Before the DM physics analysis in the next subsection, we consider the color triplet fermions included in the ${\bf 10}$-plets. 
Although they are unstable, their lifetime can be very long 
  since they decay through the colored Higgs boson and the GUT gauge boson which are very heavy. 
If the colored particles decay after Big Bang Nucleosynthesis (BBN) with the age of the universe around 1 second, 
  the energetic decay products could destroy light nuclei which have been successfully synthesized during BBN. 
We can simply avoid this problem if the lifetime of the colored fermion is shorter than 1 second. 
In this subsection, we discuss how to realize this situation.

In Sec.~\ref{sec:GCUandPD}, we have investigated gauge coupling unification by introducing the mass splitting between the doublet and the triplet components in ${\bf 10}$-plet fermions ($10_{E}^{(1,2)}$). 
We have found that this mass splitting results in gauge coupling unification below the Planck scale, $M_{GUT} < M_P$. 
This mass splitting is also important to shorten the color triplet fermions lifetime. 
We can generate the mass splitting by employing the Dimopolouos-Wilczek mechanism \cite{DWms1}. 
Consider Yukawa interaction for $10_{E}^{(1,2)}$ fermions with the ${45}_H$ in Eq.~(\ref{eq:ExoticY}). 
Following Ref.~\cite{DWms1}, 
    we set the VEV for ${45}_H$ in the $B-L$ direction: 
    $\langle { 45}_H\rangle = M_I \times diag (1, 1, 1 , 0 , 0) \times i \sigma_2$ \cite{DWms1}. 
Thus, the mass terms for the doublets and the triplet components of the ${\bf 10}$-plets are expressed as 
\bea
 {\cal L}_{\rm mass} \supset
\begin{pmatrix} 
{\bar D}^{(1)} & {\bar D}^{(2)} \end{pmatrix}   
\begin{pmatrix} 
m_{10}^{(1)}  & 0 
\\ 0 & m_{10}^{(2)} 
\end{pmatrix} 
\begin{pmatrix} D^{(1)}  \\ D^{(2)}  \end{pmatrix} 
+ 
\begin{pmatrix} {\bar T}^{(1)}  & {\bar T}^{(2)}  \end{pmatrix}   
\begin{pmatrix} 
m_{10}^{(1)}  & m_{45}
\\  m_{45} & m_{10}^{(2)}
\end{pmatrix} 
\begin{pmatrix} T^{(1)}  \\ T^{(2)}  \end{pmatrix}, 
\label{eq:DWmass} 
\eea
where $m_{10}^{(1,2)} = Y_A^{(1,2)} M_I $  and $m_{45} = Y_{45}^{(12)}M_I$. 
As in the previous section, we set $m_{10}^{(1)} = m_{10}^{(2)} \equiv m_D$, and the mass eigenvalues of the triplet fermions are $m_{T}^{(1,2)} = |m_D \pm m_{45}|$. 
Setting $m_{45} = m_T \gg m_D$, we obtain almost degenerate triplet fermions masses, 
  $m_{T}^{(1,2)} \simeq m_T \gg m_D$. 
This is the setup in the previous section.

Let us now estimate the lifetime of the color triplet fermions.  
A triplet fermion decays into a doublet fermion in ${\bf 10}$-plet  and the SM quark and lepton through an off-shell GUT gauge boson 
  $({\bf 6, 2, 2}) \supset {\bf 45}$ in the PS gauge group decomposition. 
The partial lifetime of this process is calculated to be 
\bea
\tau_{T} &\simeq& 192 \pi^3 \frac{M_{GUT}^4}{m_{T}^5}.  
\label{eq:tautrip}  
\eea
From Fig.~\ref{fig:MIandMGUT}, the proton lifetime constraint yields an upper bound 
  on the triplet fermion mass for fixed $m_D$ values. 
Eq.~(\ref{eq:tautrip}) implies that the minimum lifetime of the triplets is determined by the upper bound on $m_T$. 
We find $m_D \lesssim 2$ TeV to satisfy the BBN constraint, $\tau_T<1$ s  
  for a corresponding maximum value of $m_T$. 
A triplet fermion also decays into a $SO(10)$ singlet fermion,  
  top quark and tau lepton through an off-shell colored Higgs boson. 
The partial lifetime of this process is calculated to be 
\bea
\tau_{T} &\simeq& \frac{192 \pi^3}{Y_{t}^2 {Y_H}^2} \frac{m_{HC}^4}{m_{T}^5} \simeq 1 \; {\rm s} \left(\frac{m_{HC}[{\rm GeV}]}{3.0 \times 10^{14}}\right)^4 \left(\frac{5.0 \times 10^4 }{m_{T}[{\rm TeV}]}\right)^5 \left(\frac{55}{m_0 [{\rm GeV}]}\right)^2, 
\label{eq:lifetime3E}
\eea
where $Y_{t} \simeq 1$ is the SM top Yukawa coupling, and we express $Y_H$ in terms of a new parameter $m_0$ defined as ${Y_H} = \left(\sqrt{2} m_0/v_h\right)$. 
This new parameter plays an important role in the DM physics analysis in the next sub-section as well as in the analysis in Sec.~\ref{sec:HiggsStab}. 
For our benchmark values used in the following sections, $m_{T} = 5.0\times 10^4$ GeV and $m_0 = 55$ GeV, 
the BBN constraint of $\tau_T <1$ s to an upper bound on the colored Higgs boson mass. 
Combining with the lower bound on the colored Higgs boson mass from the proton lifetime constraint, 
  we find 
\bea
4.5 \times 10^{11}< m_{HC} [{\rm GeV}] < 3.0 \times 10^{14}. 
\label{eq:mHCbound}
\eea
As we have mentioned in the previous section, our results for the gauge coupling unification remain almost the same even for $m_{HC}$ values in this range.

\subsection{Singet-Doublet Fermion Dark Matter}
\label{sec:SDDM}
In our model, the DM candidate is a linear combination of the $SU(2)_L$ doublets
   in the ${\bf 10}$-plet and the singlet fermions. 
The doublet and singlet fermions individually acquire their masses from the VEVs of $\Phi_A$ and $\Phi_B$ as 
\bea
 {\cal L}  \supset \sum_i \left( m_{D}^{(i)} D^{(i)} {\bar D}^{(i)} + m_S^{(i)} 1_E^{(i)} 1_E^{(i)} \right), 
\eea
where 
\bea
m_{D}^{(i)} = Y_{A}^{(i)} \langle \Phi_A\rangle  = \frac{1}{\sqrt{2}} Y_{A}^{(i)} M_I, \qquad 
m_S^{(i)} = Y_{B}^{(i)} \langle \Phi_B\rangle=  \frac{1}{\sqrt{2}} Y_{B}^{(i)} M_I. 
\eea 
In Sec.~\ref{sec:GCUandPD}, we have set $m_D^{(1)} = m_D^{(2)} = m_D = {\cal O} (1)$ TeV. 
In addition, the Yukawa interactions involving $10_H$ in Eq.~(\ref{eq:ExoticY}) generate the mixing masses between the doublets and the singlets after electroweak symmetry breaking: 
\bea
{\cal L} 
&\supset&  
\sum_{i, j} { Y_H}^{(ij)} {1}_{E}^{(i)}  { 10}_{E}^{(j)} {10}_{H} 
\nonumber \\
&\supset&  {Y_H}^{(ij)} \left(1_E^{(i)} D^{(j)} H_d+  1_E^{(i)} {\bar D}^{(j)}  H_u\right)
\nonumber \\
&\supset&{Y_H}^{(ij)} ( \cos\beta \;  1_E^{(i)} D^{(j)} {H}^\dagger+ \sin\beta \; 1_E^{(i)} {\bar D}^{(j)}  H) .    
\label{eq:YHD}
\eea
For simplicity, we choose only ${Y_H}^{(1,1)} \equiv Y_H$ to be sizable and real,
   and only consider the first generation for our DM physics discussion. 
Thus, the relevant Lagrangian is given by 
\bea
 {\cal L}  \supset m_{D} D {\bar D} + m_S S S + {Y_H} \left( \cos\beta D {H}^\dagger S + \sin\beta {\bar D} H S \right) + {\rm h.c}, 
\label{eq:YukawaDM}
\eea
where we have introduced a new notation, $D^{(1)} \equiv D$ and $1_E^{(1)} \equiv  S$. 
Substituting $H = 1/\sqrt{2} (0, h + v_h)^T$ ($h$ is the SM Higgs boson and $v_h = 246$ GeV is the Higgs VEV), 
  we obtain the mass matrix for the electrically neutral fermions: 
\bea
{\cal L}  \supset  \frac{1}{2}
\begin{pmatrix} D_0 & {\bar D_0} &  S \end{pmatrix}   
\begin{pmatrix} 
0  &  m_{D} & m_0 \sin\beta
\\ m_{D}  &  0  &  m_0  \cos\beta
\\ m_0 \sin\beta  & m_0  \cos\beta  & m_S
\end{pmatrix} 
\begin{pmatrix} 
 D_0 \\ {\bar D_0} \\  S
\end{pmatrix}, 
\label{eq:massmat} 
\eea 
where $m_0 \equiv {Y_H} v_h/\sqrt{2}$. 
This symmetric mass matrix can be diagonalized by a single orthogonal matrix $U$ 
   for the mass eigenstates $\psi^{1,2,3}$ with masses $m_{1,2,3}$ 
   defined as $( \psi_1, \psi_2, \psi_3)^{T} = U^{-1} ( D,  {\bar D} , S)^{T}$. 
The lightest mass eigenstate is identified with the DM particle.

To simplify the DM analysis, we consider two extreme cases: (i) ${ m_S \gg m_{D}}$, 
where the DM is mostly the doublet component (a linear  combination of $\psi_1$ and $\psi_2$). (ii) ${m_S \ll m_{D}}$, 
where the DM is mostly the singlet component ($\psi_{3}$). 
The first case is similar to the Higgsino-like neutralino DM scenario in the Minimal Supersymmetric SM. 
This case has been well studied in the literature (see, for example, \cite{ArkaniHamed:2006mb}), 
    where the correct DM relic density is reproduced with the DM mass of around 1 TeV.  
In the following, we will focus on case (ii). 
For $m_S, m_0 \ll m_{D}$ in this case, the mass eigenvalues can be approximated as 
\bea
m_{1,2} \simeq m_D, \qquad m_{3} \simeq  m_S - m_0 \left(\frac{m_0}{m_{D}}\right) \sin2\beta, 
\label{eq:DMmass}
\eea
From Eq.~(\ref{eq:YukawaDM}), we extract the interactions involving the DM particle ($\psi_3$),  
\bea
{\cal L}  &\supset&  \frac{1}{2}
\begin{pmatrix} \psi_1 & \psi_2 &  \psi_3 \end{pmatrix} 
U^T
\begin{pmatrix} 
0  &  0 & \frac{m_0 \sin\beta}{v_h}  h
\\ 0  &  0  &   \frac{m_0 \cos\beta}{v_h}   h
\\  \frac{m_0 \sin\beta}{v_h}  h  &  \frac{m_0 \cos\beta}{v_h}   h  & 0
\end{pmatrix} 
U
\begin{pmatrix} 
 \psi_1 \\ \psi_2 \\  \psi_3
\end{pmatrix}, 
\\ \nonumber
&\supset & \frac {1}{2} y_{33} h  \psi_{3}  \psi_{3} + \frac{1}{2} y_{31} h  \psi_{3}  \psi_{1} + \frac{1}{2} y_{32} h  \psi_{3}  \psi_{2}, 
\label{eq:Lint} 
\eea 
where the couplings $y_{31}$, $y_{32}$, and $y_{33}$ are determined by the elements of the mixing matrix $U$, $m_0$ and $\beta$.

The thermal relic density of the DM particle is evaluated by solving the Boltzmann equation, 
\bea 
  \frac{dY}{dx}
  = - \frac{\langle \sigma v \rangle}{x^2}\frac{s (m_{3})}{H(m_{3})} \left( Y^2-Y_{EQ}^2 \right), 
\label{eq:Boltzman}
\eea  
    where $x = m_{3}/T$, $H(m_{3})$ is the Hubble parameter and the yield ($Y = n/s$) is given 
    by the ratio of the DM number density ($n$) and the entropy density ($s$), 
    and $Y_{EQ}$ is the yield of the DM particle in thermal equilibrium:  
\bea 
s(m_{3}) = \frac{2  \pi^2}{45} g_\star m_{3}^3,  \; \; 
 H(m_{3}) =  \sqrt{\frac{\pi^2}{90} g_\star} \frac{m_{3}^2}{M_P}, \; \; 
 Y_{EQ}(x) =  \frac{g_{DM}}{2 \pi^2} \frac{x^2 m_{3}^3}{s(m_{3})} K_2(x),   
\eea 
with $K_2$ being the Bessel function of the second kind. 
In Eq.~(\ref{eq:Boltzman}), $\langle\sigma v\rangle$ is the thermal average 
   of the total pair annihilation cross section of the DM particles 
   times their relative velocity: 
\bea 
\langle \sigma v \rangle =  \frac{g_{DM}^2}{64 \pi^4}
  \left(\frac{m_{3}}{x}\right) \frac{1}{n_{EQ}^{2}}
  \int_{4 m_{3}^2}^\infty  ds \; 2 (s- 4 m_{3}^2) \sigma(s) \sqrt{s} K_1 \left(\frac{x \sqrt{s}}{m_{3}}\right),
\label{eq:ThAvgSigma}
\eea
where $g_{DM} = 2$ denotes the degrees of freedom of the Majorana fermion DM particle,  
$n_{EQ}=s(m_{3}) Y_{EQ}/x^3$ is the equilibrium number density of the DM particle, $K_1$ is the modified Bessel function of the first kind, and $\sigma (s)$ is the total annihilation cross section of the DM particle.  
The DM particle density at the present time is evaluated from
\bea 
  \Omega_{DM} h^2 =\frac{m_{3} s_0 Y(x\to\infty)} {\rho_c/h^2}, 
\eea 
where $\rho_c/h^2 =1.05 \times 10^{-5}$ GeV/cm$^3$  is the critical density, and $s_0 = 2890$ cm$^{-3}$ is the entropy density of the present Universe.

In order to evaluate $\sigma (s)$, we consider two processes for the pair annihilation of the DM particles: 
   the $t/u$-channel processes mediated by the $\psi_{1,2}$ or charged fermions in $D$ and ${\overline D}$, 
   and the $s$-channel process mediated by the SM Higgs boson. 
For the $t/u$-channel processes with $m_D \gg m_3$, 
    we consider the effective Lagrangian after integrating out $\psi_{1,2}$, 
\bea
 {\cal L}_{eff} = \frac{1}{2} \left(\frac{y_{31}^2}{m_D}\right) h h \psi_{3} \psi_{3} + \frac{1}{2} \left(\frac{y_{32}^2}{m_D}\right) h h \psi_{3} \psi_{3}. 
\eea 
For example, the cross section for $\psi_{1}$ mediated processes is estimated as 
\bea
\sigma_0  \simeq \left(\frac{1}{64\pi}\right) \left(\frac{y_{31}^2}{m_{D}}\right)^2 \simeq   y_{31}^4  \left( \frac{1 {\rm TeV}}{m_{D}} \right)^2 {\rm pb}. 
\eea 
Here, we have assumed $m_3 > m_h$. 
Since the DM is mostly the singlet component, its coupling with the SM Higgs boson is suppressed, $y_{31}^4 \ll1$. 
Therefore, the cross section for this process is much smaller than a typical cross section of $1$ pb for a thermal DM. 
We can apply the same discussion for $\psi_2$ and charged fermion mediated process, 
  and conclude that the cross sections for all the $t/u$-channel processes are too small to reproduce the observed DM relic density.

Let us next consider the $s$-channel process mediated by the SM Higgs boson. 
Although the DM coupling with the SM Higgs is suppressed, the $s$-channel cross section
  can be enhanced if the DM mass is close to the Higgs boson resonance point, $ m_3\simeq m_h/2$. 
For $m_3 \ll m_D$, this will turn out to be the only possibility for reproducing the observed DM relic density.  
The $s$-channel cross section is given by
\bea
{\sigma} (s) = \frac{y_{33}^2}{64} 
\left( 3 \left(\frac{m_b}{v_h} \right)^2 + 3 \left(\frac{m_c}{v_h} \right)^2+ 3 \left(\frac{m_\tau}{v_h} \right)^2  \right) \frac{\sqrt{s(s-4 m_{3}^2})}{\left(s- m_h^2\right)^2 + m_h^2 \Gamma_h^2}. 
\eea
For the final states, we have considered a pair of bottom (b) quarks, charm (c) quarks, and tau ($\tau$) leptons 
  with masses $m_b = 2.82$ GeV,  $m_c = 685$ MeV, and $m_\tau = 1.75$ GeV \cite{Bora:2012tx}, respectively.  
 $\Gamma_h = \Gamma_h^{SM}+ \Gamma_h^{DM}$ is the total decay width of the SM Higgs boson, 
  where $\Gamma_h^{SM} = 4.07$ MeV \cite{Denner:2011mq} is the SM Higgs boson decay width in the SM and 
\bea
\Gamma_h^{DM} = \theta\left(m_h-2 m_3\right)\frac{y_{33}^2}{16\pi}m_h\left(1-\frac{4m_3^2}{m_h^2}\right)^{3/2}
\eea
    is the partial decay width of the SM Higgs boson decay into a pair of DM particles. 
The annihilation cross section is determined by only two free parameters, $m_3$ and $y_{33}$.
After numerically solving the Boltzmann equation with the $s$-channel cross section, 
   we find the relation between $m_3$ and $y_{33}$ to reproduce the observed DM relic density  
   of $\Omega_{DM}h^2 = 0.120$ \cite{Aghanim:2018eyx}.

\subsection{Direct Detection Bound on Dark Matter}
\label{sec:DD}

\begin{figure}[t]
\begin{center}
\includegraphics[scale =0.91]{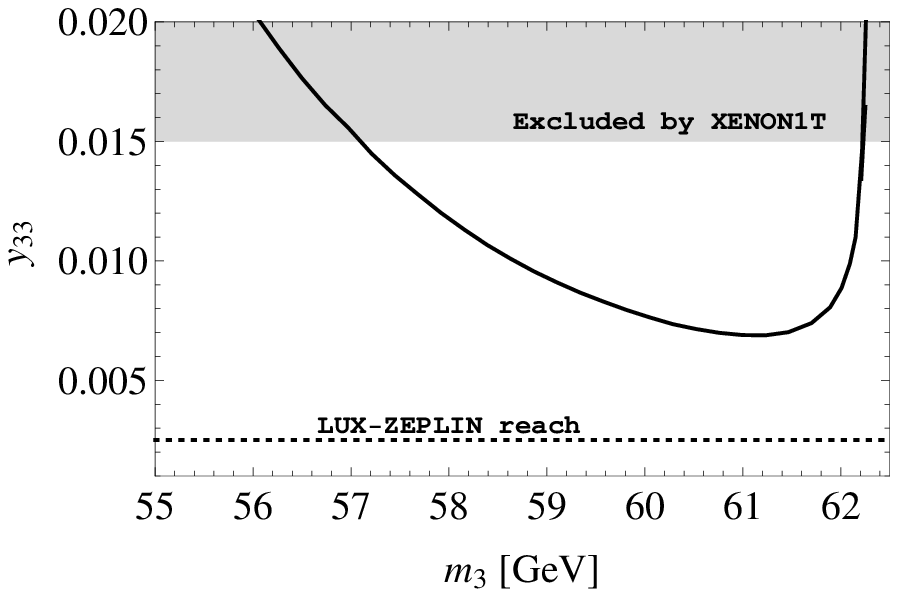}\;\;
\includegraphics[scale=0.85]{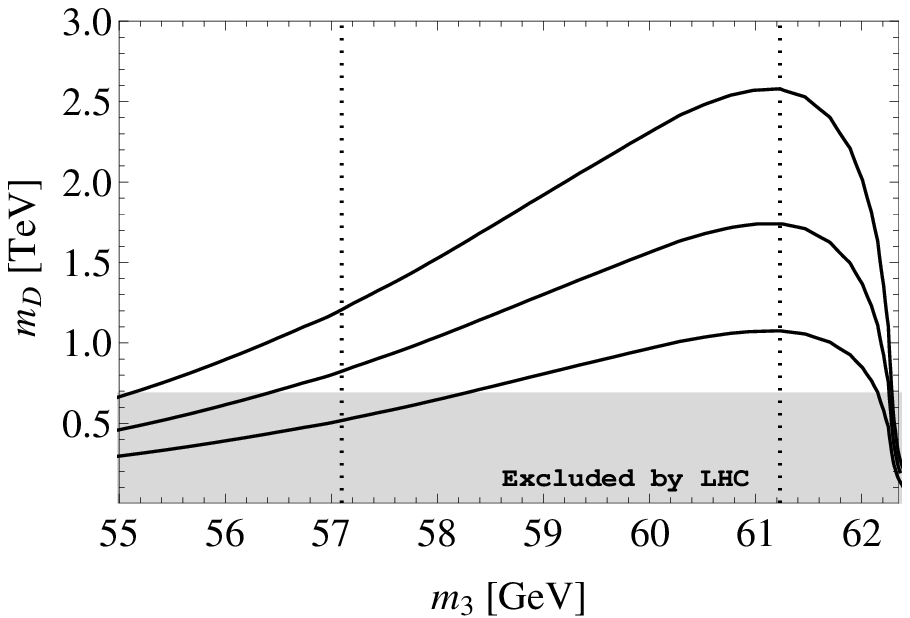}       
\end{center}
\caption{
Left panel: $y_{33}$ as a function of  $m_{3}$ (solid curve) along which the observed DM density  is reproduced. 
The gray shaded region is excluded by the XENON1T results and the horizontal dashed line marks 
  the search reach of the future LUX-ZEPLIN experiment. 
Right panel: $m_D$ as a function of $m_3$ for different choices of $m_0 [{\rm GeV}] = 55, 45$ and $35$ 
  (solid curves from top to bottom) and fixed $\beta= \pi/3$, 
  corresponding to the solid curve in the left panel.  
The gray shaded region is excluded by the null LHC search results for a heavy charged lepton. 
}
\label{fig:DM}
\end{figure}

Various experiments to directly search for the DM particles are in operation. 
The most severe constraint on the so-called spin-independent (SI) cross section 
  of the DM particle scattering off nuclei is given by the XENON1T direct DM detection experiment \cite{Aprile:2018dbl}. 
We use this result to constrain the parameter space for $m_3$ and $y_{33}$. 
The SI elastic cross section for the DM scattering off a nucleon is given by  
\bea
\sigma_{\rm SI} \simeq \frac{1}{\pi} \left(\frac{y_{33}}{v_h}\right)^2  \left(\frac{\mu_{\rm eff}}{m_h^2}\right)^2 f_N^2, 
\label{eq:SIcs1}
\eea
where $\mu_{\rm eff} = m_N m_{3}/(m_N +m_{3})$ is the effective mass of the DM-nucleon system 
      with a nucleon mass $m_N = 0.939$ GeV \cite{Patrignani:2016xqp}. 
The nuclear matrix element of a nucleon $f_N$ is given by 
\bea
f_N = \left(\sum_{q = u,d,s} f_{T_q} + \frac{2}{9} f_{T_G}\right) m_N,
\label{eq:NME}
\eea
where $f_{T_q}$ values are determined by lattice QCD analysis:  
  $f_{T_u} +f_{T_d} \simeq 0.056$ \cite{Ohki:2008ff} for up (u) and down (d) quarks 
  and $\left|f_{T_s}\right| \leq 0.08$ \cite{Ohki:2008ff} for strange (s) quark, 
  and $f_{T_G}$ is determined using trace anomaly condition, $\sum_{q = u,d,s} f_{T_q} + f_{T_G} = 1$ \cite{QCDanomaly}. 
Using a conservative value for $f_{T_s} =0$, we obtain $f_N^2 \simeq 0.0706 m_N^2$ 
  and the SI cross section is given by 
\bea
\sigma_{\rm SI} \simeq 4.47 \times 10^{-7} {\rm pb} \times y_{33}^2, 
\label{eq:SIcs2}
\eea  
where we have used $m_{3} \gg m_N$.

In the left panel of Fig.~\ref{fig:DM}, we plot $y_{33}$ as a function of $m_{3}$ (solid black curve) 
   along which the observed DM relic density, $\Omega_{DM}h^2 = 0.120$, is reproduced.  
For $m_{3} \simeq m_h/2$, XENON1T constraint, $\sigma_{\rm SI} \leq 1.0\times 10^{-10}$ pb \cite{Aprile:2018dbl}, leads to an upper bound on $y_{33} \leq 1.50 \times 10^{-2}$ from Eq.~(\ref{eq:SIcs2}). 
The gray shaded region is excluded by the XENON1T, 
and the allowed region for the DM mass lies in the range 
of $57.10 \lesssim m_{3}[{\rm GeV}] \lesssim 61.23$. 
The next generation LUX-ZEPLIN (LZ) experiment will improve the cross section bound significantly, 
$\sigma_{\rm SI} \leq 2.8\times 10^{-12}$  pb \cite{Akerib:2018lyp}, 
which corresponds to $y_{33} \leq 2.51 \times 10^{-3}$. 
This search reach is depicted as  the horizontal dashed line.     
We can see that the current allowed region will be fully covered by the LZ experiment.

Both $m_3$ and $y_{33}$ are determined in terms of the model parameters $\beta$, $m_0$, $m_S$, and $m_D$. 
As shown in the left panel of Fig.~\ref{fig:DM}, 
   $y_{33}$ is determined as a function of $m_{3}$ 
   in order to reproduce the observed DM relic density. 
Hence, $m_D$ is determined as a function of $m_0$, $\beta$, and $m_3$. 
In the right panel of Fig.~\ref{fig:DM}, we plot $m_D$ as a function of $m_3$ 
   for different choices of $m_0 [{\rm GeV}] = 55, 45$ and $35$ (solid curves from top to bottom) and fixed $\beta= \pi/3$. 
The allowed range of the DM mass of $57.10 \lesssim m_{3}[{\rm GeV}] \lesssim 61.23$ 
   is indicated by the two vertical dotted lines,  
   which bound the allowed mass range of $m_D$ for different $m_0$ values.  
The gray shaded region is excluded by a lower mass bound of $m_D < 690$ GeV 
   from the CMS search result for a heavy charged lepton at the LHC \cite{CMS:2018cgi}.

\section{Stability of the SM Higgs Potential}
\label{sec:HiggsStab}

Because of the large top Yukawa coupling, the SM Higgs quartic coupling turns negative 
   around $\mu \simeq 10^{10}$ GeV  \cite{Buttazzo:2013uya}. 
This implies that the electroweak vacuum of the SM is unstable, which is, in principle, known as the Higgs potential instability problem. 
It may not be a serious problem for the SM because the electroweak vacuum is meta-stable with lifetime much longer than the age of the universe. 
However, in the GUT scenario, the SM Higgs is embedded inside a GUT Higgs multiplet 
   and the negative quartic coupling of the SM Higgs may imply that some of the GUT Higgs multiplets 
   have negative quartic couplings which can make the GUT vacuum unstable.  
To avoid this problem, we impose the condition that the SM Higgs quartic coupling remains positive up to the PS SSB scale.

Let us evaluate the RG evolution of the SM Higgs quartic coupling ($\lambda_H$), 
   to which the new ${\bf 10}$-plets fermions also contribute, in addition to the SM particles. 
As discussed in Sec.~\ref{sec:GCUandPD}, the ${\bf 10}$-plets modify the RG running of the SM gauge couplings,  
   which in turn modifies the RG running of $\lambda_H$. 
In addition, the doublets in the ${\bf 10}$-plet fermions contribute to the beta-function of $\lambda_H$ 
   through their Yukawa couplings with the SM Higgs doublets in Eq.~(\ref{eq:YukawaDM}). 
The RG equation of $\lambda_H$ is expressed as 
\bea
\mu  \frac{d \lambda_H}{d \mu} &=&  \frac{1}{16\pi^2} \left(\beta_{SM} +\theta(\mu - m_{D}) \left(4 \lambda_H  {Y_H}^2\ -  4 {Y_H}^4 \right)\right), 
\label{eq:HiggsRGE}  
\eea
where ${Y_H} = \left(\sqrt{2} m_0/v_h\right)$, and $\beta_{SM}$ denotes the beta function of the SM. 
The contribution of the doublet fermions (the second term in the right-hand side of Eq.~(\ref{eq:HiggsRGE})) 
   is analogous to the top quark contribution, $\beta_{SM} \supset 12 y_t^2 \lambda_H - 12 y_t^4$, 
   where $y_t$ is the top-quark coupling. 
The presence of such a coupling is effectively equivalent to the SM with a larger $y_t$. 
Hence, the Yukawa coupling may make the situation worse and destabilize the Higgs potential 
   at an energy scale even lower than $\mu \simeq 10^{10}$ GeV. 
However, the presence of ${\bf 10}$-plet fermions also modify the running of the SM gauge couplings 
   which generates a positive contribution to $\beta_{SM}$. 
See, for example, Ref.~\cite{Gogoladze:2010in}, where the authors have shown 
   that the Higgs potential stability problem can be solved in the presence of TeV scale new fermions.  
We now show that the instability problem can also be solved in our model in the presence of the ${\bf 10}$-plet fermions.

\begin{figure}[t]
\begin{center}
\includegraphics[scale =0.8]{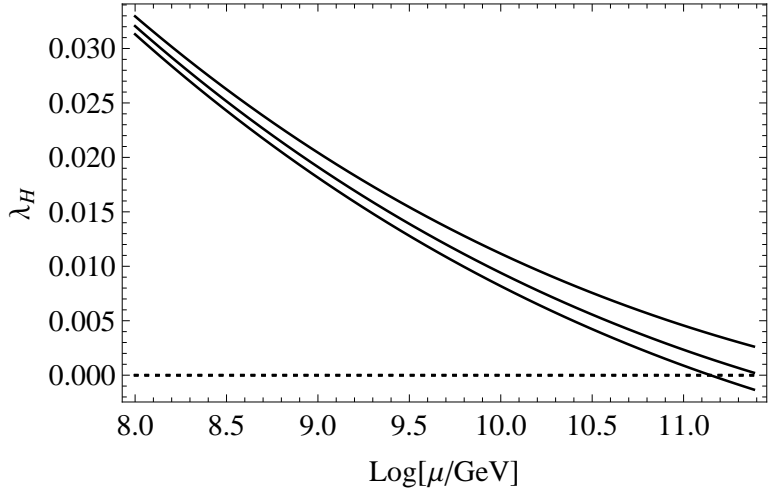} 
\end{center}
\caption{
RG running of $\lambda_H$ for $\mu < M_I \simeq 2 \times 10^{11}$ GeV, 
  for $m_{D} = 2$ TeV and $m_{T} = 5\times 10^4$ TeV. 
The solid lines from top to bottom correspond to $m_0 [{\rm GeV}]=35$, $55$, and $60$, respectively. 
The dotted line depicts $\lambda_H = 0$. 
}
\label{fig:HStab}
\end{figure}

The RG running of $\lambda_H$ is determined by three parameters: $m_{D}$, $m_{T}$ and ${m_0}$. 
In the following analysis, we approximate $Y_H$ to be a constant. 
For fixed values of $m_{D}$, $m_{T}$ and ${m_0}$, we numerically solve the RG equations. 
In Fig.~\ref{fig:HStab}, we show the RG running of $\lambda_H$ 
   for $m_{D} = 2$ TeV, $m_{T} = 5 \times 10^4$ TeV, and  $m_0 [{\rm GeV}]=35$, $55$, and $60$ (solid lines from top to bottom).  
The horizontal dotted line represents $\lambda_H = 0$. 
From the gauge coupling unification analysis in Sec.~\ref{sec:GCUandPD}, 
   we have found  $M_I \simeq 2\times 10^{11}$ GeV for $m_{D} = 2$ TeV and $m_{T} = 5 \times 10^4$ TeV. 
In order to keep $\lambda_H (\mu) >0$ for $\mu< M_I$, we have found an upper bound on $m_0 \lesssim 55$ GeV. 
We have checked that the running of $Y_H$ can be ignored to a good approximation 
  for $m_0 \lesssim 55$ GeV or, equivalently, $Y_H \lesssim 0.32$. 

\section{Conclusion}
\label{sec:conc}

We have proposed a simple non-supersymmetric GUT model 
   based on the gauge group $SO(10) \times U(1)_\psi$.    
The model includes three generations of fermions 
   in ${\bf 16}$ ($+1$), ${\bf 10}$ ($-2$) and ${\bf 1}$ ($+4$) representations. 
In addition to the ${\bf 16}$-plets that contains the SM fermions plus RHNs, 
the ${\bf 10}$-plet and singlet fermions are introduced. 
In the presence of the new fermions, the model is free from all the gauge and mixed gauge-gravitational anomalies. 
With the new fermions and a suitable set of Higgs fields, 
gauge coupling unification is achieved in two-step breaking of  $SO(10)$ to the SM. 
Namely, the SM gauge couplings are partially unified in the PS group at the intermediate scale of $M_{I} = 10^{12}-10^{11}$ GeV with the PS group subsequently  unified into $SO(10)$ group at $M_{GUT} = 5\times 10^{15}-10^{16}$ GeV. 
Since the Majorana masses for the RHNs are generated through the PS symmetry breaking, 
successful gauge coupling unification leads to the natural scale for the seesaw mechanism. 
We have found a correlation between $M_{GUT}$ and $M_{I}$, namely $M_{I}$ is increases as $M_{GUT}$ is decreases. 
Hence, the proton lifetime is predicted to be shorter for a higher $M_{I}$ value, 
   which can be tested by the Hyper-Kamiokande experiment in the future.

The new ${\bf 10}$-plet and singlet fermions have Yukawa couplings 
  with two $SO(10)$-singlet $U(1)_\psi$ Higgs fields and 
  the fermion masses are generated once the $U(1)_\psi$ symmetry is broken 
  by the $U(1)_\psi$ Higgs fields VEVs.  
The $U(1)_\psi$ Higgs filed $\Phi_A$ which has the Yukawa coupling with the ${\bf 10}$-plet fermions 
  is identified with the inflaton. 
We have shown that through its gauge and Yukawa interactions, 
  the effective inflaton potential exhibits an approximate inflection-point 
  and successful inflection-point inflation is realized. 
The Hubble parameter during the inflation is found to be much smaller than the PS symmetry breaking scale, $H_{inf} < M_I$, so that the cosmologically unwanted monopoles 
  generated by the breaking of the GUT and the PS symmetries are diluted away.  
With a suitable choice of the model parameters, 
  the reheating temperature after inflation can be high enough for a successful thermal leptogenesis 
  while low enough not to restore the PS gauge symmetry.

With the Higgs field contents of our model, a ${\bf Z}_2$ symmetry remains unbroken after the GUT symmetry breaking,  
   and the lightest Majorana mass eigenstate from linear combinations of the ${\bf 10}$-plets and singlet fermions 
    is stable and thus a viable DM candidate of our model. 
We focus on the case that the DM particle is mostly composed of the $SO(10)$ singlet fermion 
    and it communicates with the SM particles through the Higgs-portal interactions. 
For this Higgs-portal fermion DM scenario,     
    we have identified the model parameter region to reproduce the observed DM relic density 
    while satisfying the current constraint from the direct DM detection experiments. 
The present allowed region will be fully covered by the future direct detection experiments such as LZ experiment. 
Finally, we have shown that in the presence of the new fermions, 
    the SM Higgs potential is stabilized up to $M_I$.

\section*{Acknowledgements}
N.O. would like to thank the Particle Theory Group of the University of Delaware for hospitality during his visit.
This work of is supported in part by the United States Department of Energy Grants DE-SC0012447 (N.O) 
  and DE-SC0013880 (D.R and Q.S) and Bartol Research Grant BART-462114 (D.R).



\begin{thebibliography}{}

\bibitem{GUT}
  J.~C.~Pati and A.~Salam,
  Phys.\ Rev.\ Lett.\  {\bf 31}, 661 (1973).
  J.~C.~Pati and A.~Salam,
  Phys.\ Rev.\ D {\bf 10}, 275 (1974)
  Erratum: [Phys.\ Rev.\ D {\bf 11}, 703 (1975)];
  H.~Georgi, H.~R.~Quinn and S.~Weinberg,
  Phys.\ Rev.\ Lett.\  {\bf 33}, 451 (1974).



\bibitem{SUSYGUT}
  S.~Dimopoulos, S.~Raby and F.~Wilczek,
  Phys.\ Rev.\ D {\bf 24}, 1681 (1981);
  L.~E.~Ibanez and G.~G.~Ross,
  Phys.\ Lett.\  {\bf 105B}, 439 (1981);
  M.~B.~Einhorn and D.~R.~T.~Jones,
  Nucl.\ Phys.\ B {\bf 196}, 475 (1982);
  T.~E.~Clark, T.~K.~Kuo and N.~Nakagawa,
  Phys.\ Lett.\  {\bf 115B}, 26 (1982);
  W.~J.~Marciano and G.~Senjanovic,
  Phys.\ Rev.\ D {\bf 25}, 3092 (1982);
  P.~Langacker and M.~x.~Luo,
  Phys.\ Rev.\ D {\bf 44}, 817 (1991);
  C.~Giunti, C.~W.~Kim and U.~W.~Lee,
  Mod.\ Phys.\ Lett.\ A {\bf 6}, 1745 (1991);
  U.~Amaldi, W.~de Boer and H.~Furstenau,
  Phys.\ Lett.\ B {\bf 260}, 447 (1991).



\bibitem{nonSUSYGUT}
  H.~Georgi and S.~L.~Glashow,
  Phys.\ Rev.\ Lett.\  {\bf 32}, 438 (1974). 



\bibitem{S010nonSUSY1} 
  G.~Lazarides, Q.~Shafi and C.~Wetterich,
  Nucl.\ Phys.\ B {\bf 181}, 287 (1981);



\bibitem{S010nonSUSY}
  R.~Holman, G.~Lazarides and Q.~Shafi,
  Phys.\ Rev.\ D {\bf 27}, 995 (1983).
  B.~Bajc, A.~Melfo, G.~Senjanovic and F.~Vissani,
  Phys.\ Rev.\ D {\bf 73}, 055001 (2006)
  [hep-ph/0510139];
  S.~Bertolini, L.~Di Luzio and M.~Malinsky,
  Phys.\ Rev.\ D {\bf 80}, 015013 (2009)
  [arXiv:0903.4049 [hep-ph]];
  A.~S.~Joshipura and K.~M.~Patel,
  Phys.\ Rev.\ D {\bf 83}, 095002 (2011)
  [arXiv:1102.5148 [hep-ph]];
  L.~Di Luzio,
  arXiv:1110.3210 [hep-ph];
  S.~Bertolini, L.~Di Luzio and M.~Malinsky,
  Phys.\ Rev.\ D {\bf 85}, 095014 (2012)
  [arXiv:1202.0807 [hep-ph]];
  S.~Bertolini, L.~Di Luzio and M.~Malinsky,
  AIP Conf.\ Proc.\  {\bf 1467}, 37 (2012)
  [arXiv:1205.5637 [hep-ph]];
  G.~Altarelli and D.~Meloni,
  JHEP {\bf 1308}, 021 (2013)
  [arXiv:1305.1001 [hep-ph]]; 
  K.~S.~Babu and S.~Khan,
  Phys.\ Rev.\ D {\bf 92}, no. 7, 075018 (2015)
  [arXiv:1507.06712 [hep-ph]]. 



\bibitem{Seesaw} 
P.~minkowski, Phys. Lett. B {\bf 67}, 421 (1977);
T.~Yanagida, in {\it Proceedings of the Workshop on the Unified
  Theory and the Baryon Number in the Universe} (O.~Sawada and
  A.~Sugamoto, eds.), KEK, Tsukuba, Japan, 1979, p.~95;
M.~Gell-Mann, P.~Ramond, and R.~Slansky, {\it Supergravity} (P.~van
  Nieuwenhuizen et al. eds.), North Holland, Amsterdam, 1979, p.~315;
S.~L. Glashow, {\it The future of elementary particle physics}, in
  {\it Proceedings of the 1979 Carg{\`e}se Summer Institute
 on Quarks and Leptons} (M.~L{\'e}vy et al. eds.),
 Plenum Press, New York, 1980, p.~687;
R.~N. Mohapatra and G.~Senjanovi{\'c},
  ``Neutrino Mass and Spontaneous Parity Violation,''
  Phys.\ Rev.\ Lett.\  {\bf 44}, 912 (1980); 
J.~Schechter and J.~W.~F.~Valle,
  ``Neutrino Masses in SU(2) x U(1) Theories,''
  Phys.\ Rev.\ D {\bf 22}, 2227 (1980).    




\bibitem{leptogenesis1}
  M.~Fukugita and T.~Yanagida,
  Phys.\ Lett.\ B {\bf 174}, 45 (1986).
  S.~Davidson, E.~Nardi and Y.~Nir,
  Phys.\ Rept.\  {\bf 466}, 105 (2008)
  [arXiv:0802.2962 [hep-ph]]. 






\bibitem{Tdefects}
  A.~Vilenkin,
  Phys.\ Rept.\  {\bf 121}, 263 (1985);
  R.~Jeannerot, J.~Rocher and M.~Sakellariadou,
  Phys.\ Rev.\ D {\bf 68}, 103514 (2003)
  [hep-ph/0308134];
  J.~Chakrabortty, R.~Maji, S.~K.~Patra, T.~Srivastava and S.~Mohanty,
  Phys.\ Rev.\ D {\bf 97}, no. 9, 095010 (2018)
  [arXiv:1711.11391 [hep-ph]];
  J.~Chakrabortty, R.~Maji and S.~F.~King,
  Phys.\ Rev.\ D {\bf 99}, no. 9, 095008 (2019)
  [arXiv:1901.05867 [hep-ph]].





\bibitem{monopole1}
  G.~'t Hooft,
  Nucl.\ Phys.\ B {\bf 79}, 276 (1974);
  A.~M.~Polyakov,
  JETP Lett.\  {\bf 20}, 194 (1974)
  [Pisma Zh.\ Eksp.\ Teor.\ Fiz.\  {\bf 20}, 430 (1974)];
  P.~Langacker and S.~Y.~Pi,
  Phys.\ Rev.\ Lett.\  {\bf 45}, 1 (1980).


\bibitem{monopole2}
  G.~Lazarides, M.~Magg and Q.~Shafi,
  Phys.\ Lett.\  {\bf 97B}, 87 (1980);
  G.~Lazarides and Q.~Shafi,
  Phys.\ Lett.\  {\bf 148B}, 35 (1984).



\bibitem{monopole3}
  L.~Patrizii and M.~Spurio,
  Ann.\ Rev.\ Nucl.\ Part.\ Sci.\  {\bf 65}, 279 (2015)
  [arXiv:1510.07125 [hep-ex]].




\bibitem{Guth} 
  A.~H.~Guth,
  Phys.\ Rev.\ D {\bf 23}, 347 (1981)
  [Adv.\ Ser.\ Astrophys.\ Cosmol.\  {\bf 3}, 139 (1987)].




\bibitem{Shafi:2006cs} 
  Q.~Shafi and V.~N.~Senoguz,
  Phys.\ Rev.\ D {\bf 73}, 127301 (2006)
  [astro-ph/0603830].



\bibitem{Okada:2010jf} 
  N.~Okada, M.~U.~Rehman and Q.~Shafi,
  Phys.\ Rev.\ D {\bf 82}, 043502 (2010)
  [arXiv:1005.5161 [hep-ph]].



\bibitem{Okada:2014lxa} 
  N.~Okada, V.~N.~Senoguz and Q.~Shafi,
  Turk.\ J.\ Phys.\  {\bf 40}, no. 2, 150 (2016)
  [arXiv:1403.6403 [hep-ph]].



\bibitem{Senoguz:2015lba} 
  V.~N.~Senoguz and Q.~Shafi,
  Phys.\ Lett.\ B {\bf 752}, 169 (2016)
  [arXiv:1510.04442 [hep-ph]].



\bibitem{Linde:1993cn} 
  A.~D.~Linde,
  Phys.\ Rev.\ D {\bf 49}, 748 (1994)
  [astro-ph/9307002].



\bibitem{IPI}
  N.~Okada and D.~Raut,
  Phys.\ Rev.\ D {\bf 95}, no. 3, 035035 (2017)
  [arXiv:1610.09362 [hep-ph]];
  N.~Okada, S.~Okada and D.~Raut,
  Phys.\ Rev.\ D {\bf 95}, no. 5, 055030 (2017)
  [arXiv:1702.02938 [hep-ph]].
 


\bibitem{leptogenesis2}
  S.~Davidson and A.~Ibarra,
  Phys.\ Lett.\ B {\bf 535}, 25 (2002)
  [hep-ph/0202239];
  W.~Buchmuller, P.~Di Bari and M.~Plumacher,
  Annals Phys.\  {\bf 315}, 305 (2005)
  [hep-ph/0401240];
  F.~Buccella, D.~Falcone, C.~S.~Fong, E.~Nardi and G.~Ricciardi,
  Phys.\ Rev.\ D {\bf 86}, 035012 (2012)
  [arXiv:1203.0829 [hep-ph]]. 

\bibitem{SO10Z2}
  T.~W.~B.~Kibble, G.~Lazarides and Q.~Shafi,
  Phys.\ Rev.\ D {\bf 26}, 435 (1982). 




\bibitem{Ferrari:2018rey} 
  S.~Ferrari, T.~Hambye, J.~Heeck and M.~H.~G.~Tytgat,
  Phys.\ Rev.\ D {\bf 99}, no. 5, 055032 (2019)
  [arXiv:1811.07910 [hep-ph]].





\bibitem{Planck2018}
  Y.~Akrami {\it et al.} [Planck Collaboration],
  arXiv:1807.06211 [astro-ph.CO].



\bibitem{RunningSpectral} 
  K.~N.~Abazajian {\it et al.},
  ``Inflation Physics from the Cosmic microwave Background and Large Scale Structure,''
  Astropart.\ Phys.\  {\bf 63}, 55 (2015)
  [arXiv:1309.5381 [astro-ph.CO]].



\bibitem{Buttazzo:2013uya} 
  D.~Buttazzo, G.~Degrassi, P.~P.~Giardino, G.~F.~Giudice, F.~Sala, A.~Salvio and A.~Strumia,
  JHEP {\bf 1312}, 089 (2013)
  [arXiv:1307.3536 [hep-ph]].



\bibitem{Nath:2006ut} 
  P.~Nath and P.~Fileviez Perez,
  Phys.\ Rept.\  {\bf 441}, 191 (2007)
  [hep-ph/0601023].



\bibitem{Miura:2016krn} 
  K.~Abe {\it et al.} [Super-Kamiokande Collaboration],
  Phys.\ Rev.\ D {\bf 95}, no. 1, 012004 (2017)
  [arXiv:1610.03597 [hep-ex]].



\bibitem{Abe:2011ts} 
  K.~Abe {\it et al.},
  arXiv:1109.3262 [hep-ex].



\bibitem{SDDM}
  T.~Cohen, J.~Kearney, A.~Pierce and D.~Tucker-Smith,
  Phys.\ Rev.\ D {\bf 85}, 075003 (2012)
  [arXiv:1109.2604 [hep-ph]]; 
  C.~Cheung and D.~Sanford,
  JCAP {\bf 1402}, 011 (2014)
  [arXiv:1311.5896 [hep-ph]];
  L.~Calibbi, A.~Mariotti and P.~Tziveloglou,
  JHEP {\bf 1510}, 116 (2015)
  [arXiv:1505.03867 [hep-ph]];
  S.~M.~Boucenna, M.~B.~Krauss and E.~Nardi,
  Phys.\ Lett.\ B {\bf 755}, 168 (2016)
  [arXiv:1511.02524 [hep-ph]];
  N.~Maru, T.~Miyaji, N.~Okada and S.~Okada,
  JHEP {\bf 1707}, 048 (2017)
  [arXiv:1704.04621 [hep-ph]].





%


\bibitem{DWms1}
  S.~Dimopoulos and F.~Wilczek,
  Print-81-0600 (SANTA BARBARA), NSF-ITP-82-07;
  [hep-ph/9306242].



\bibitem{ArkaniHamed:2006mb} 
  N.~Arkani-Hamed, A.~Delgado and G.~F.~Giudice,
  Nucl.\ Phys.\ B {\bf 741}, 108 (2006)
  [hep-ph/0601041].



\bibitem{Bora:2012tx} 
  K.~Bora,
  Horizon {\bf 2}, 112 (2013)
  [arXiv:1206.5909 [hep-ph]].



\bibitem{Denner:2011mq} 
  A.~Denner, S.~Heinemeyer, I.~Puljak, D.~Rebuzzi and M.~Spira,
  Eur.\ Phys.\ J.\ C {\bf 71}, 1753 (2011)
  [arXiv:1107.5909 [hep-ph]].



\bibitem{Aghanim:2018eyx} 
  N.~Aghanim {\it et al.} [Planck Collaboration],
  arXiv:1807.06209 [astro-ph.CO].



\bibitem{Aprile:2018dbl} 
  E.~Aprile {\it et al.} [XENON Collaboration],
  Phys.\ Rev.\ Lett.\  {\bf 121}, no. 11, 111302 (2018)
  [arXiv:1805.12562 [astro-ph.CO]].



\bibitem{Patrignani:2016xqp} 
  C.~Patrignani {\it et al.} [Particle Data Group],
  Chin.\ Phys.\ C {\bf 40}, no. 10, 100001 (2016).



\bibitem{Ohki:2008ff} 
  H.~Ohki {\it et al.},
  Phys.\ Rev.\ D {\bf 78}, 054502 (2008)
  [arXiv:0806.4744 [hep-lat]].



\bibitem{QCDanomaly}
  R.~J.~Crewther,
  Phys.\ Rev.\ Lett.\  {\bf 28}, 1421 (1972);
  M.~S.~Chanowitz and J.~R.~Ellis,
  Phys.\ Lett.\  {\bf 40B}, 397 (1972);
  M.~S.~Chanowitz and J.~R.~Ellis,
  Phys.\ Rev.\ D {\bf 7}, 2490 (1973);
  J.~C.~Collins, A.~Duncan and S.~D.~Joglekar,
  Phys.\ Rev.\ D {\bf 16}, 438 (1977);
  M.~A.~Shifman, A.~I.~Vainshtein and V.~I.~Zakharov,
  Phys.\ Lett.\  {\bf 78B}, 443 (1978).



\bibitem{Akerib:2018lyp} 
  D.~S.~Akerib {\it et al.} [LUX-ZEPLIN Collaboration],
  arXiv:1802.06039 [astro-ph.IM].



\bibitem{CMS:2018cgi} 
  CMS Collaboration [CMS Collaboration],
  CMS-PAS-EXO-18-005.


\bibitem{Gogoladze:2010in} 
  I.~Gogoladze, B.~He and Q.~Shafi,
  Phys.\ Lett.\ B {\bf 690}, 495 (2010)
  [arXiv:1004.4217 [hep-ph]]; 
%
  N.~Okada and Q.~Shafi,
  Phys.\ Lett.\ B {\bf 747}, 223 (2015)
  [arXiv:1501.05375 [hep-ph]];
%
H.~Y.~Chen, I.~Gogoladze, S.~Hu, T.~Li and L.~Wu,
  Eur.\ Phys.\ J.\ C {\bf 78}, no. 1, 26 (2018)
  [arXiv:1703.07542 [hep-ph]];
  N.~Okada, S.~Okada and D.~Raut,
  arXiv:1811.11927 [hep-ph].

  
\end{thebibliography}
\end{document}